%% file: z8ps_3dhst_apj_v2.tex
\documentclass[twocolumn,astrosym,tighten,apj]{aastex63}
\usepackage{graphicx,amssymb,amsmath,lineno}    
\usepackage{amsthm,amsfonts,amscd,wasysym,cleveref}

\received{---}
\revised{---}
\accepted{---}
\submitjournal{\apj}

\shorttitle{$z\sim7$-$8$ point sources in \HSTsurvey\ Fields}
\shortauthors{Ishikawa et al.}
\graphicspath{{./}{}}

\newcommand\sex{\texttt{SExtractor}}
\newcommand\eazy{\texttt{EAZY}}
\newcommand\bags{\texttt{Bagpipes}}
\newcommand\emcee{\texttt{emcee}}
\newcommand\lmfit{\texttt{LMFIT}}
\newcommand\astropy{\texttt{astropy}}

\newcommand\candels{CANDELS}
\newcommand\HSTsurvey{3D-HST}

\newcommand{\hst}{\textit{HST}}

\newcommand{\spitz}{\textit{Spitzer}}
\newcommand{\sborg}{{SuperBoRG}}

\newcommand{\oiii}{[\textrm{O}~\textsc{iii}]}

\newcommand{\civ}{[\textrm{C}~\textsc{iv}]}

\newcommand{\ly}{${\rm Ly\alpha}$}

\newcommand{\hb}{${\rm H\beta}$}

\newcommand{\simgt}{\,\rlap{\lower 3.5 pt \hbox{$\mathchar \sim$}} \raise
1pt \hbox {$>$}\,}
\newcommand{\simlt}{\,\rlap{\lower 3.5 pt \hbox{$\mathchar \sim$}} \raise
1pt \hbox {$<$}\,}

\DeclareMathOperator\erf{erf}
\DeclareMathOperator\ulim{ulim}

\newcommand{\ii}{$I_{814}$}
\newcommand{\yy}{$Y_{105}$}
\newcommand{\jj}{$J_{125}$}
\newcommand{\hh}{$H_{160}$}

\newcommand{\aminsq}{arcmin$^2$}
\newcommand{\zph}{$z_{ph}$}
\newcommand{\muv}{$M_{UV}$}
\newcommand{\buv}{$\beta_{UV}$}
\newcommand{\veff}{$V_{\rm eff}$}
\newcommand{\myr}{$M_{\astrosun}$ yr$^{-1}$}

\begin{document}

\title{Unresolved $z\sim8$ point sources and their impact on the bright end of the galaxy luminosity function}

\newcommand{\jhu}{The William C. Miller III Department of Physics and Astronomy, Johns Hopkins University, Baltimore, MD 21218, USA}
\newcommand{\stsci}{Space Telescope Science Institute, 3700 San Martin Drive, Baltimore, MD 21218, USA}
\newcommand{\melborne}{School of Physics, Tin Alley, University of Melbourne, VIC 3010, Australia}

\correspondingauthor{Yuzo Ishikawa}
\email{yishika2@jhu.edu}

\author[0000-0001-7572-5231]{Yuzo Ishikawa}
\affiliation{\jhu}
\affiliation{\stsci}


\author{Takahiro Morishita}
\affiliation{\stsci}
\affiliation{IPAC, California Institute of Technology, MC 314-6, 1200 E. California Boulevard, Pasadena, CA 91125, USA}

\author{Massimo Stiavelli}
\affiliation{\stsci}

\author{Nicha Leethochawalit}
\affiliation{\melborne}
\affiliation{ARC Centre of Excellence for All Sky Astrophysics in 3 Dimensions (ASTRO 3D), Australia}
\affiliation{National Astronomical Research Institute of Thailand (NARIT), MaeRim, Chiang Mai, 50180, Thailand}

\author{Harry Ferguson}
\affiliation{\stsci}

\author{Roberto Gilli}
\affiliation{Osservatorio di Astrofisica e Scienza dello Spazio di Bologna, Via P. Gobetti 93/3, 40129 Bologna, Italy}

\author{Charlotte Mason}
\affiliation{Cosmic Dawn Center (DAWN)}
\affiliation{Niels Bohr Institute, University of Copenhagen, Jagtvej 128, 2200 København N, Denmark}


\author{Michele Trenti}
\affiliation{\melborne}

\author{Tommaso Treu}
\affiliation{Department of Physics and Astronomy, University of California, Los Angeles, Los Angeles, CA 90095, USA}

\author{Colin Norman}
\affiliation{\jhu}
\affiliation{\stsci}




\begin{abstract}
The distribution and properties of the first galaxies and quasars are critical pieces of the puzzle in understanding galaxy evolution and cosmic reionization. 
Previous studies have often excluded unresolved sources as potential low redshift interlopers. We combine broadband color and photometric redshift analysis with morphological selections to identify a robust sample of candidates consistent with unresolved point sources at redshift $z\sim8$ using deep \textit{Hubble Space Telescope} (\hst) images. 
We also examine G141 grism spectroscopic data to identify and eliminate dwarf star contaminants. From these analyses, we identify three, bright ($M_{UV}\simlt-22$\,ABmag) dropout point sources at $7.5<z<8.1$. 
Spectral energy distribution analyses suggest that these sources are either quasars or compact star-forming galaxies. The flux captured by the IRAC 4.5\,$\mu$m channel suggests that they have moderate \hb$+$\oiii\ equivalent widths. We calculate the number density of point sources at $z\sim7$-8, and find that a double powerlaw model well describes the point source distribution. 
We then extend our analysis to estimate the combined point source + galaxy luminosity function and 
find that the point sources have a non-negligible contribution to the bright-end excess. 
The fact that the point sources dominate only at $M_{UV}\simlt-22$ suggests that their contribution to cosmic reionization is likely limited. 
While spectroscopic follow-up is needed to confirm the nature of these point sources, this work demonstrates that the inclusion of Lyman dropout point sources is necessary for a complete census of the early galaxies at the epoch of cosmic reionization.  
\end{abstract}

\keywords{cosmology: reionization --- galaxies: high-redshift --- galaxies: luminosity function --- galaxies: photometry --- galaxies: formation --- quasars: general}


\section{Introduction} \label{sec:intro}

Statistical studies of the first galaxies and quasars are crucial to understanding their formation and evolution processes. To-date, a tremendous amount of effort has been made to probe the early universe with high redshift surveys like CANDELS \citep{Koekemoer2011ApJS197, Grogin2011ApJS197, Bouwens2019ApJ880}, BoRG \citep{Trenti2011ApJ, Bradley2012ApJ, Morishita2018ApJ, morishita2021sb}, HUDF12 \citep{HUDF12}, XDF \citep{XDF}, CLASH \citep{CLASH}, HFF \citep{HFF}, RELICS \citep{RELICS}, ULTRAVISTA \citep{McCracken2012AA, Stefanon2017ApJ,Stefanon2019ApJ,Bowler2020MNRAS}, among others. These surveys combined with follow-up spectroscopy have successfully identified some of the earliest galaxies up to $z\sim9$\,-\,10, yet characterizing the number densities and physical properties of these early sources remain incomplete. 
It is necessary to accurately quantify the early populations with observational constraints. 

Characterizing the luminosity function is a fundamental step in estimating the contribution from various luminous sources; it describes the number density of sources as a function of luminosity, or absolute magnitude. Since ultraviolet (UV) emission is primarily dominated by ionizing sources, the rest-frame UV luminosity function is a useful tool in investigating the early galaxy populations. In particular, the shape of the luminosity function can provide insights into the different physical processes such as star-formation and quasar activity that drive galaxy formation. 
The faint end of the luminosity function is believed to be the key driver for cosmic reionization \citep[e.g.,][]{Ishigaki2018ApJ, atek2018MNRAS}, in which the early universe transitioned from completely neutral to almost ionized \citep{Ouchi2010ApJ723, konno14, pentericci14, Robertson2015ApJ802L, Mason2018ApJ857L,Mason2019MNRAS485,Hoag2019ApJ}. It is believed that reionization paved the way for the formation of the first galaxies \citep{LoebBarkana2001ARAA}, yet the question of what astrophysical objects are primarily responsible for reionization remain debated. 

The bright end of the luminosity function is composed of the brightest sources that may be signposts of in-situ star formation or even quasar activity. The discovery of luminous quasars \citep{Mortlock2011nat,Banados2018Nature, Yang2020ApJ, Wang2021z76} and luminous star forming galaxies at $z\simgt7$ \citep{zitrin15,Oesch16,Hashimoto2018Natur557,jiang21} may indicative of populations of luminous sources that are unaccounted for. There is no consensus on the shape of the bright end luminosity function; some even suggest that the early galaxy luminosity function may depart from the standard Schechter form \citep{Harikane2022ApJS}. This departure manifest as a bright end excess \citep[e.g.,][]{Morishita2018ApJ}, and its origins remain unclear. Theoretical studies suggest that this bright excess may be caused by intense and compact star-forming clumps \citep[e.g.,][]{Ma2018MNRAS} or even stochastic quasar activity \citep[e.g.,][]{Ren2020ApJ}. Luminous source are also believed to contribute to cosmic reionization to some degree, yet the consensus on their contribution remain controversial  \citep[e.g.,][]{willott2010,Finkelstein2015ApJ,Jiang2016ApJ, Matsuoka2019ApJ883, naidu2020ApJ}. The characterization of the brightest sources at high redshifts remain elusive.  

High-redshift sources are typically identified with the Lyman dropout photometric selection \citep{Steidel1996AJ} combined with follow-up spectroscopic confirmation. However, the complication is that the redshifted spectral energy distribution (SED) of these sources at the end of cosmic reionization, $z\sim7$-8, overlaps with those of low-mass foreground stars. As a result, previous studies have often excluded compact, unresolved sources with star-like morphology in preference for more galaxy-like sources with extended morphology. Nevertheless, there is evidence from lensing surveys that early galaxies are very compact \citep[e.g.,][]{Bouwens2017ApJ, salmon2020ApJ}. Some even predict compact star-forming clumps \citep{Ma2018MNRAS}. So there is a possibility that certain population of quasars and compact galaxies at at $z\sim7$-8 that are rejected with the standard selection. 

There is a renewed interest in examining these overlooked point sources. A recent medium-depth, wide $\sim0.4 \textrm{ deg}^2$ \hst\ survey (\sborg; \citealt{Morishita2020z8, morishita2021sb}) has identified several $z\gtrsim8$ point sources as potential quasar candidates. Key features of these $z\sim8$ point sources are their blue rest-frame UV slopes and the \spitz/IRAC flux excess in the SED, which may be indicative of significant \hb\ and \oiii\ emission often seen in quasars. Observations suggest that these point sources are unlikely to be foreground stars and may contribute to the bright end of the luminosity function. 



In this study, as part of our \hst\ archival program (AR 15804; PI. Morishita), we reexamine the selection of high-redshift compact unresolved (point) sources that have been overlooked in previous studies. We take advantage of the successful \sborg\ study to revisit $z\sim7$-8 point sources in the \candels\ legacy \hst\ fields. Since the selection criteria of our study and of \sborg\ are complementary, we combine the results of both studies to characterize the $z\sim7$-8 dropout point sources and to quantify their contribution to the total galaxy luminosity function. For simplicity, we will refer to these sources as ``point sources'' throughout the paper.


This paper is organized as follows. In Section \ref{sec:data} we describe the data reduction and target selection from \HSTsurvey. In Section \ref{sec:res} we explore the properties of the targets selected. And in Section \ref{sec:disc} we discuss the physical implications of these objects. We use the AB-magnitude system \citep{OkeGunn1983ApJ,fukugita96} and adopt the $h=0.7$, $\Omega_M=0.3$, and $\Omega_{\Lambda}=0.7$ cosmology.

\section{Target Selection} \label{sec:data}

\subsection{Source Catalog}\label{sec:select}
Our primary focus is to identify sources that satisfy the dropout color selection and have point source morphology in the \candels\ fields. We begin our analyses with the publicly available photometric catalogs provided by the \HSTsurvey\ team \citep{Brammer2012ApJS200,vandokkum13}. The \HSTsurvey\ is a \hst\ near-infrared spectroscopic survey designed to study galaxies across the universe. It surveyed nearly 700 \aminsq\ of the well-studied \hst/\candels\ Treasury fields to obtain direct images and spectroscopic data with the ACS/G800L and WFC3/G141 grisms. \HSTsurvey\ covers about 75\% of the original \candels\ area. When all the fields are combined, their photometric observations of \hh\ reach median $5\,\sigma$ depths at 26 mags at $1\,\arcsec$ aperture. Further details of the survey and the published catalog can be found in \cite{Skelton2014ApJS214, Momcheva2016ApJS225}. 

Our choice of using the \HSTsurvey\ catalog over the catalogs published by the \candels\ team \citep{guo13,Galametz13,Stefanon17,Nayyeri17,Barro2019ApJS243} is that uniform analysis is performed on all 5 \candels\ fields by the \HSTsurvey\ team to create the source catalog. This vastly simplifies the source detection procedure (Sec.~\ref{subsec:color}) and the completeness simulation analysis (Sec.~\ref{sec:LF}), to calculate the number density of the target population. However, due to inconsistencies in the filter coverage, we only analyze 4 of the 5 \candels\ fields (AEGIS, COSMOS, GOODS South, and UKIDSS-UDS), where F814W, F125W, and F160W filters are available (Sec.~\ref{subsec:color}). 
We also exploit the published G141 grism data to identify low redshift interlopers (Sec.~\ref{subsec:lowz}).

We obtain deep \hst\ data from the publicly available \HSTsurvey\ database. The \hst\ image mosaics used have already been corrected for distortions and drizzled to the plate scale of $0.\!\arcsec06 \textrm{ pixel}^{-1}$. The photometric source catalogs were produced using point-spread function (PSF)-matched aperture photometry, reduced using \sex\ \citep{Bertin1996}, and flux calibrated to an aperture radius of $0.\!\arcsec7$. The \hst\ ACS and WFC3 images were convolved to match the \hst/F160W PSF ($\sim0.\!\arcsec14$). Ground-based optical, NIR, and \spitz/IRAC fluxes are similarly PSF-matched to a combination of F125W, F140W, and F160W priors and aperture corrected to F160W (or F140W, otherwise). 

\subsection{Color-dropout and shape selection}\label{subsec:color}
Our strategy in identifying $z\sim7$-8 point source candidates from the \HSTsurvey\ is twofold. First, we identify sources with the Lyman-break dropout technique \citep{Steidel1996AJ} from the photometric catalog. Then, we select point sources from the list of Lyman-dropout sources. The color selection is only based on deep \hst\ photometry. 
A caveat is that unlike the \cite{Morishita2020z8, morishita2021sb} selection, which uses the F105W/F125W/F160W (\yy/\jj/\hh) filters, the \HSTsurvey\ catalog does not include \yy\ fluxes. Instead, we use the \cite{Bouwens2015ApJ} color-dropout criteria, which is based on the F814W/F125W/F160W (\ii/\jj/\hh) selection:

\begin{figure}
	 \begin{center}
    	 \begin{tabular}{c}
    	 \includegraphics[width=0.97\columnwidth]{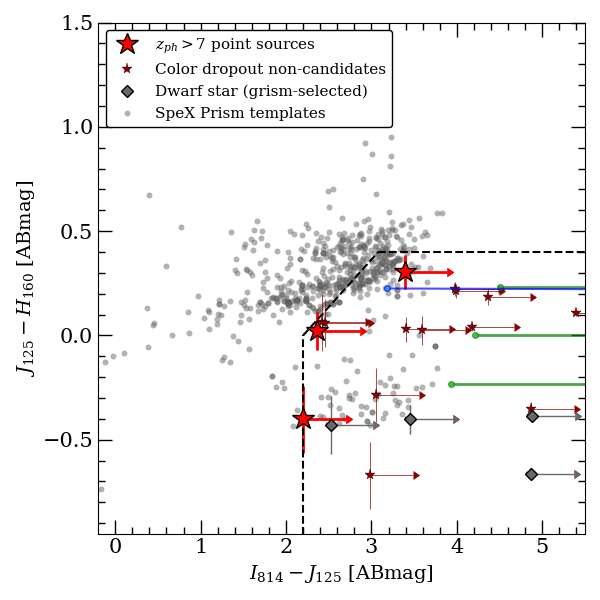}
    	 \end{tabular}
	 \end{center}
	 \caption{Color-color diagram for the target selections from the \HSTsurvey. The targets fall into the marked by the dashed lines that define the \cite{Bouwens2015ApJ} color selection. Our final point sources from Table \ref{tab:targlist} are shown as red stars. Point sources that do not meet the photometric redshift selection are shown as dark red stars. Spectroscopically confirmed dwarf stars are shown as gray diamonds. We also compare with dwarf star template colors \citep{spexprism2014} in gray dots, predicted quasar colors from composite SDSS quasar spectrum (\citealt{sdssQ14}) in blue, and simple powerlaw spectra at $\beta=-1,-2,-3$ in green. Strong overlap of known dwarf star colors and our Lyman-dropout sources indicate the need for further follow-up to disentangle the degeneracy.}
	 \label{fig:colorcolor} 
\end{figure}

\begin{equation}
    \begin{split}
    S/N_{125,160} > 5.0 \\
    S/N_{\textrm{blue}} < 2.0 \\
    I_{814} - J_{125}  > 2.2 \\
	J_{125} - H_{160}  < 0.4 \\
	I_{814} - J_{125}  > 2\times(J_{125} - H_{160})+2.2
    \end{split}
\end{equation}

\input{targlist}

\begin{figure*}[!hbt]
	 \begin{center}
    	 \begin{tabular}{c}
    	 \includegraphics[width=0.98\textwidth]{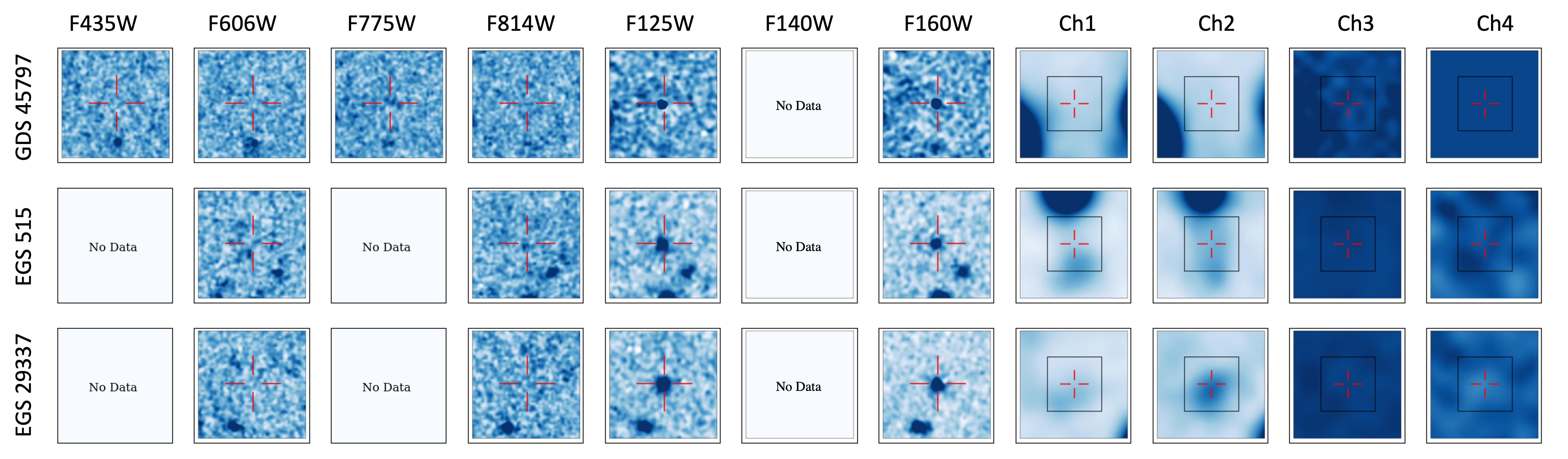}
    	 \end{tabular}
	 \end{center}
	 \caption{
	 \hst\ ACS and WFC3 3 arcsec$^2$ postage stamp images of the point source targets shown at 95 percentile fluxes.  \spitz/IRAC images are shown in 6~arcsec$^2$ boxes with 3 arcsec$^2$ box inserts in black. Filter images with no observations are indicated as ``No Data.'' Since the IRAC Ch3 and CH4 fluxes are dominated by noise, their fluxes are not included in the analysis.}
	 \label{fig:imgTOT} 
\end{figure*}

Compared to the \citet{Morishita2020z8} selection, this color selection results in a broader $z\sim7$-8 selection. Also, the GOODS-North catalog does not include \ii\ data, so we only examine 4 of the 5 \candels\ fields. Where available, we use the \hst/ACS blue filters (\ii\ and bluer) to determine the Lyman-dropout with strict blue signal-to-noise constraints, combined with the non-detection flag from the \hst\ pipeline catalog. Since we require $2\,\sigma$ non-detections for $I_{814}$ fluxes, we calculate the resulting $I_{814} - J_{125}$ lower limit color as:
\begin{equation}
(I_{814} - J_{125})_{\textrm{lim}} = -2.5\log_{10}(2\sigma_{814} / J_{125}).
\end{equation}
We do not include sources with $I_{814} - J_{125}$ that fall below the non-detection limit (i.e. falls outside the selection box in Figure \ref{fig:colorcolor}). While additional ground-based fluxes are available, higher $S/N$ \hst\ fluxes are prioritized for color selection here (but see Sec.~\ref{subsec:lowz}).

From the color dropouts, we identify point sources based on two morphological parameters, elongation and flux concentration, measured in the \hh\ filter. The selection criteria are defined by \cite{Morishita2020z8}:
\begin{equation}
    \begin{split}
    e<1.2 \\
	f_{5}/f_{10}>0.5. 
    \end{split}
\end{equation}

\input{targphotom}

Elongation, $e$ (ratio of semi-major/semi-minor axes), describes the circularity of the source, and the flux concentration (flux ratio between inner and outer radii) describes the compactness. \cite{morishita2021sb} find light concentration is an appropriate metric for point source selection. We obtain $e$ from the \HSTsurvey\ catalog. 
To calculate flux concentration, we run \sex\ on the image mosaics, matching the \HSTsurvey\ detection parameters, and obtain detailed aperture photometry of the targets. We extract the aperture fluxes to calculate the flux ratios. 
After careful comparison of the different \hh\ flux ratios at different radii, we determine the $f_{5}/f_{10}$ flux concentration, the flux ratios taken within the 5 pixel ($0.\!\arcsec3$) and 10 pixel ($0.\!\arcsec6$) radii, to be the appropriate criteria. This decision is based on the ability to concurrently recover known dwarf star contaminants due to the point source selection (see Sec.~\ref{subsec:lowz}). Although the \HSTsurvey\ source catalogs include the \texttt{star\_class} flag that classify whether a source is star-like, \citet{Finkelstein2015ApJ} and \citet{morishita2021sb} have demonstrated that these flags are not complete down to fainter magnitudes; it fails to distinguish between fuzzy circular objects and compact point sources in the faint magnitude ranges up to $\sim24$ mag. We note that other studies \citep[e.g.,][]{rborsani2016ApJ, Bouwens2015ApJ} have successfully identified sources by combining color dropout, stellarity parameters, and SED properties. Our aim is to identify additional sources that may be missed with the standard method. 

Of the 169,614 objects listed in the \HSTsurvey\ catalog, we identify 22 \ii/\jj/\hh\ dropout point sources. Of these point sources, 7 meet the photometric redshift selection of $z_{ph}>7$, discussed in Section \ref{sec:photz}. We show these $J_{125} - H_{160}$ vs. $I_{814} - J_{125}$ color-color diagram of all color and $z_{ph}>7$ selected point sources in Figure \ref{fig:colorcolor}. Then, we check the grism spectra to further eliminate any low redshift interlopers, discussed in Section \ref{subsec:lowz}.  Our final $z\sim7$-8 point source candidates list consists of 3 point sources, which are listed in Table \ref{tab:targlist}. We also visually inspect the \hst\ images of the \zph\ selected targets to eliminate image artifacts and/or other spurious detections; postage stamp images of the final sample are shown in Figure \ref{fig:imgTOT} and discussed in Section \ref{sec:disc}. The observed fluxes of the point sources are shown in Table \ref{tab:photom}. We also list low confidence, $I_{814} - J_{125}$ limited, non-candidate point sources in Appendix \ref{apx:sourceX}.

\subsection{Photometric redshifts and SED fits}\label{sec:photz}
The Lyman-break color selection is comprehensive but also allows low-redshift sources with similar colors to migrate into the selection window. To filter out these contaminants, we apply further selection based on the photometric redshift measurement discussed here. 

\input{BAGpriors}

Following the similar approach by \cite{rborsani2021superB}, we estimate the photometric redshift, \zph, using the photometric redshift code, \eazy\ \citep{Brammer2008ApJ686}. 
While photometric redshifts are also included in the public \HSTsurvey\ catalog, they were calculated with a maximum limiting redshift of $z=6$. Hence, we re-reduce the redshift estimates for all of our color-selected samples.
We use \eazy\ in the default setup (v1.3 templates) to derive the best-fit SED and redshift posterior probability, $p(z_{ph})$. To ensure a more accurate redshift derivation, we use all available photometric data points, including ground-based fluxes that were excluded in our initial color selection in  Sec.~\ref{subsec:color}. We turn off magnitude priors in the fit to avoid any biased redshift selections, and fit between $0\leq z_{ph}\leq9$. The \zph\ fit range is based on the redshift probability from the survey completeness (Sec.~\ref{sec:LF}). From this analysis, we eliminate low redshift interlopers by making a cut in which the probability of $z_{ph}>6$ is greater than 70\%. The \zph\ and corresponding probability of our targets are shown in Table \ref{tab:targlist}. 

\input{BAGbestfit}

\begin{figure*}
	 \begin{center}
	 \includegraphics[width=\textwidth]{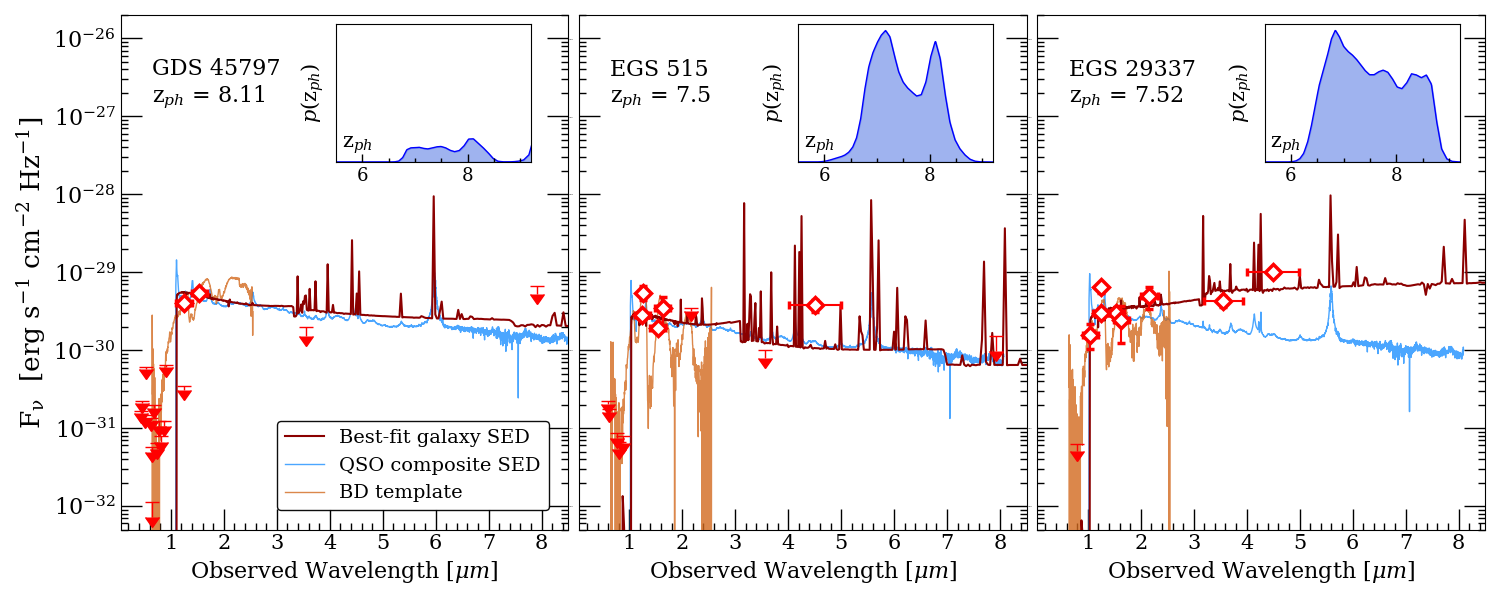}
	 \end{center}
	 \caption{Best-fit \bags\ galaxy SEDs (black) and the $1\sigma$ distribution (light gray) based on the broadband photometric data (red diamonds) with $1\sigma$ uncertainties. The x-axis errorbars indicate the filter bandwidth. We note that the we enforce $5\sigma$ cuts on \hst\ \jj\ and \hh\ fluxes. The large J-band fluxes, relative to \jj\ fluxes, in EGS 515 and EGS 29337 correspond to shallower ground-based CFHT data \citep{Bielby2012AA}. The \zph\ are fixed based on the best \eazy\ estimates. The non-detections are indicated with $2\sigma$ upper limits. We plot the best-fit quasar template SEDs (light blue). Quasars and starbursts are nearly degenerate at the predicted \zph. We can see that a galaxy model is most appropriate for EGS 29337, as expected from \cite{rborsani2016ApJ,Stark2017MNRAS}. We also plot the best-fit dwarf star templates in dark gray, which also show nearly degenerate fits. The insert plot shows the \eazy\ \zph\ probability distribution in blue. The resulting $p(z_{ph})$ distribution across $6.5\lesssim z\lesssim8.5$ is broad since \yy\ data was not included.}
	 \label{fig:eazybagsSED} 
\end{figure*}

Upon determining \zph, we refine the SED fits using \bags\ \citep{Carnall2018} to determine their physical properties.  \cite{Morishita2020z8} notes that distinguishing between luminous galaxies and quasars at $z\sim7$-8 is ambiguous and challenging without spectroscopy. In Figure \ref{fig:eazybagsSED} we plot both the best-fit quasar (described in Appendix \ref{apx:qsotemp}) and star-forming galaxy SEDs (from \bags\ fitting described next), which clearly show degenerate profiles, with the exception of EGS 29337 (to be discussed later). Since precise modeling is beyond the scope of this study, in this paper we instead assume that the sources are well represented with a young stellar spectrum with nebular emission  with \bags\ modeling and explore the inferred properties.

We describe our \bags\ fit methodology here. The redshift is fixed to \zph\ from \eazy, and we freely fit for other properties. The model fit priors are listed in Table \ref{tab:BAGpriors}, following the treatment in \cite{rborsani2021superB} as a guide. The best-fit \bags\ SEDs and \eazy\ \zph\ probability distributions are shown in Figure \ref{fig:eazybagsSED}. For each source, we use the best fitting SED model to extract the source's rest-frame UV luminosity, \muv, and other stellar parameters of interest for subsequent analysis, which is discussed in Section \ref{sec:res}. The best-fit \bags\ model parameters are shown in Table \ref{tab:bagsSED}.

\subsection{Excluding low-redshift contaminants}\label{subsec:lowz}
To further exclude low redshift interlopers among the selected point sources, we utilize G141 grism spectra made available by the \HSTsurvey\ team. As alluded to earlier, dwarf star SEDs have a sharp $1\,\micron$ drop-off that mocks the Lyman break of $z\sim7$-8 objects, making them likely interloper contaminants. There are notable spectral features at $1\,\micron$, $1.25\,\micron$, and $1.6\,\micron$ that are captured by the G141 grism. 
We find that some of our point sources are not listed in the grism catalog, due either to the extraction limit ($JH_{140}\approx26$\,mag) or incomplete spectral coverage (landing on/outside of detector edge). It is noted that the G141 grism does not cover the redshifted Ly$\alpha$ break at $z\sim7$-8, making the confirmation as high redshift sources difficult. Therefore, the objective of our inspection here is to exclude interlopers through the detection of continuum spectral features instead of characterizing their spectra.

When extracted grism spectra are available for the sources selected in Sec.~\ref{subsec:color}, we perform spectral fits to low-mass L and T dwarf template spectra, which were observed with the SpeX spectrograph on NASA InfraRed Telescope Facility \citep{Rayner2003PASP}. Template spectra are obtained from the SpeX Prism Library \citep{spexprism2014}. Based the spectral fits, we identified 4 T dwarfs with clear spectral features. Their 1D spectra are shown in Appendix \ref{apx:BDlist}. In fact, 3 of these dwarf stars were also identified in a recent \HSTsurvey\ dwarf star study \citep{Aganze2021}. We exclude these targets from the final point source list. We note that none of our final redshift selected, point source candidates were identified by \cite{Aganze2021}. However, the \cite{Aganze2021} selection was limited to spectra with $S/N>10$, and thus their selected sources are all brighter than our final targets ($H_{160}\simlt24$\,mag). This may simply reflect the limitations of the \HSTsurvey\ grism data instead of differences in selection. We also fit the photometric SEDs with SpeX dwarf star templates using \eazy. In Table \ref{tab:targlist} we list the $\chi_{\nu,BD}$, and in Figure \ref{fig:eazybagsSED} we show the best-fit models. When compared to the $\chi_{\nu}$ of \zph\ fits, we find that our final point sources are better constrained as $z\sim7$-8 sources. 

\subsection{Visual inspection of point sources}
As the final step of our sample selection, here we examine the images to identify and to eliminate any spurious fluxes in \hst\ and \spitz. Once we eliminate false detections, we repeat and refine the \eazy\ \zph\ estimates and \bags\ SED fits. 

The postage stamp images of our final point sources in Table \ref{fig:imgTOT}. Images are extracted for available deep \hst\ \citep{Grogin2011ApJS197,Koekemoer2011ApJS197,Skelton2014ApJS214} and \spitz\ \citep{Dickinson2003,Ashby2013ApJ} that make up the \HSTsurvey\ catalogs. From the images, we clearly see the blue color dropouts, which are also reflected in the SED fits. 
Some sources show suspicious blue detections. For example, the GDS 29369 images show suspicious \ii\ fluxes, despite meeting the non-detection criteria. After comparing the different filter images, we concluded that this is likely noise artifacts because their flux centroids do not match and the size is on the same order as the surrounding noise structure. The catalog also suggested spurious ground-based blue fluxes observed with \textit{Subaru} \citep{taniguchi2007ApJS} for GDS 45797. However, careful examination of the \textit{Subaru} images suggested that they are artifacts due to diffraction spikes from a nearby star. So, we treat these blue fluxes as non-detections in our analysis. Fortunately, the inclusion of these blue fluxes did not have a major affect on the \zph\ or the SED fits results. Another uncertainty comes from \spitz/IRAC fluxes, which suffer from lower spatial resolution. The \HSTsurvey\ catalog includes IRAC contamination flags for each channel; however, the flags are based on contamination within 3\!\arcsec apertures, whereas our fluxes are taken at 0.\!\arcsec7 apertures. As a result, we individually inspected the IRAC images to decide whether or not to include them in our analysis. If we visually confirmed obvious contamination within a channel, we omitted its flux. 


\begin{figure*}[!th]
	 \begin{tabular}{cc}
	 \includegraphics[width=0.97\columnwidth]{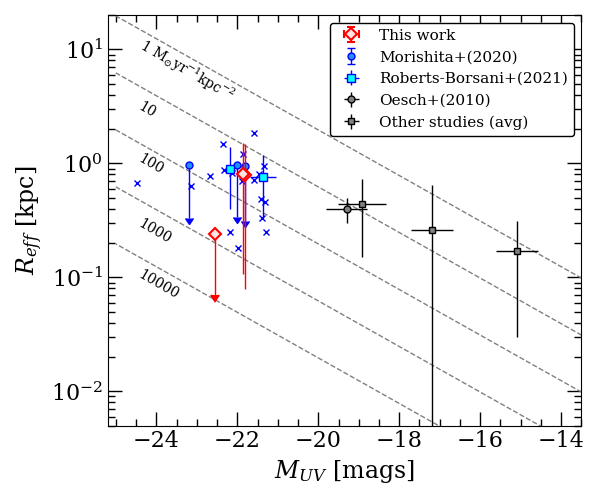} &
	 \includegraphics[width=0.97\columnwidth]{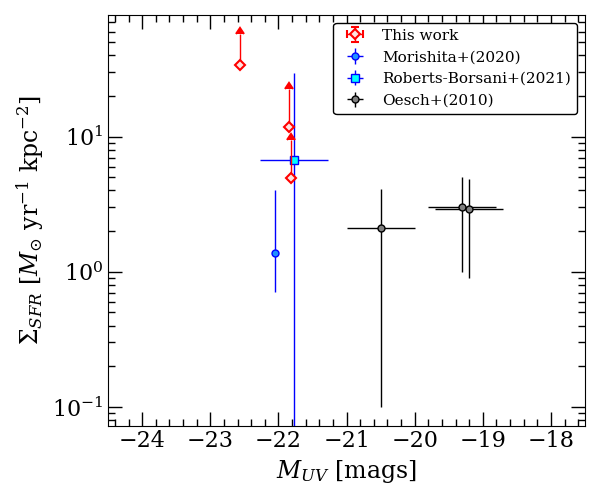} \\
	 \includegraphics[width=0.97\columnwidth]{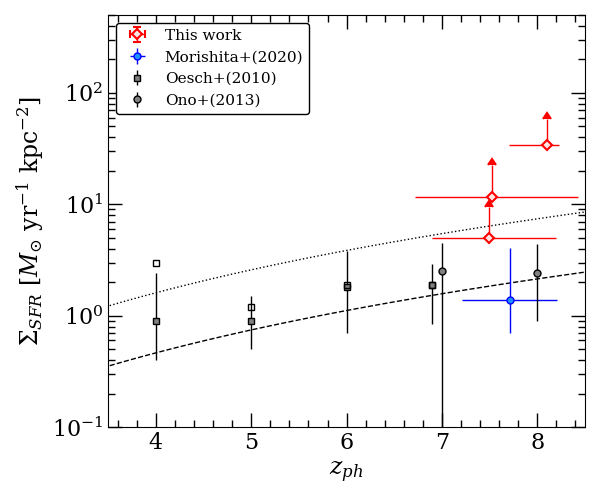} &
	 \includegraphics[width=0.97\columnwidth]{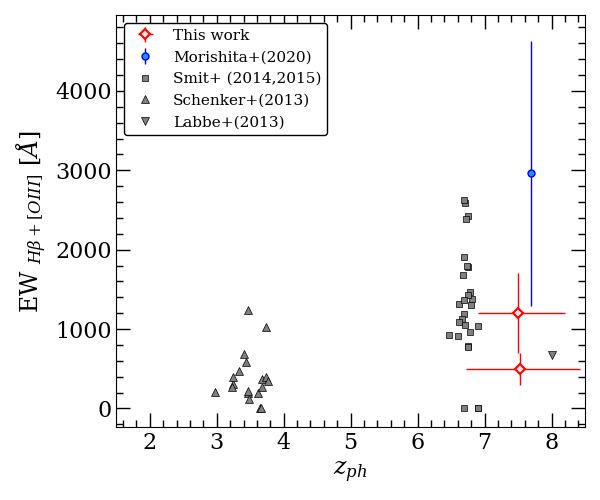} \\
	 \end{tabular}
	 \caption{
	 $R_{\textrm{eff}}$ (top left) and $\Sigma_{\textrm{SFR}}$ (top right) vs. \muv\ relations. We also plot the inferred $R_{\textrm{eff}}$-\muv\ relations from \cite{Kennicutt1998,Ono2013ApJ}.  We compare with results from \sborg\ point sources \citep{Morishita2020z8} and other galaxy studies \citep{rborsani2021superB,Oesch2010ApJ}. We also show the $\Sigma_{\textrm{SFR}}$ (bottom left) and $EW_{H\beta+\oiii}$ (bottom right) vs. redshift, and compare with \cite{Oesch2010ApJ,Ono2013ApJ, labbe2013ApJ,Schenker2013ApJ,Smit2014ApJ,Smit2015ApJ,Morishita2020z8}. We plot the inferred $\Sigma_{\textrm{SFR}}$ redshift evolution based on the empirically derived $R_{\textrm{eff}}\propto(1+z)^{-1.3}$ relations for $L^*_{z=3}$ (dotted line) and $(0.3-1)L^*_{z=3}$ (dashed line) from \cite{Ono2013ApJ}. The corresponding values for the \HSTsurvey\ point sources are shown in Table \ref{tab:bagsSED}.}
	 \label{fig:LFz8_sizeLum} 
\end{figure*}

\subsection{Cross-matching with \textit{Chandra} X-ray catalog}
\input{xraytable}
Finally, we also cross-matched our \hst\ selected source with the deep X-ray \textit{Chandra} catalogs for GOODS-South \citep{Luo2017ApJS} and AEGIS \citep{Nandra2015ApJS}. Significant X-ray emission would strengthen the case for quasar candidacy. 
Previous targeted observations, at shallower depths, have 
detected $z\sim7$ low-luminosity quasars \citep[e.g.,][]{banados2018ApJxray, Wang2021ApJ}; however, no clear X-ray emission was detected at $0.5-7 \textrm{ keV}$. We place upper limits on the fluxes and luminosities of our targets in Table \ref{tab:xray}. Moreover, there are also known Compton-thick $z>7$ quasar candidates with faint X-rays \citep{Fujimoto2022Nat}, so this is not entirely unusual. Detailed X-ray analyses will be left for future study.

\section{Results: Nature of point sources}\label{sec:res}
\subsection{On the point source selection}\label{subsec:res1}
We cross-match our point source selection with findings from \cite{Bouwens2015ApJ}, which examined the CANDELS, HUDF09, HUDF12, ERS, and BoRG/HIPPIES fields to estimate the galaxy UV luminosity function. Both studies appply similar Lyman-dropout color selections. The main difference is that we explicitly search for point-sources with the morphology selection defined in Sec.~\ref{subsec:color} using the $f_{5}/f_{10}$ flux ratios, while \cite{Bouwens2015ApJ} implements the stellarity parameter (e.g., \texttt{star\_class} flag), combined with SED fit photometry, to eliminate point-sources.

We find that only EGS 29337 is detected in both catalogs (EGSY-0120800269) with the separation of $\delta r=0.\!\arcsec11$, which is well within the PSF uncertainty. In fact, this object has also been spectroscopically confirmed as as a $z_{ph}=7.4$ galaxy \citep{rborsani2016ApJ,Stark2017MNRAS}. This means that some near point-like sources are in fact bona-fide galaxies. We discuss the implications in Section \ref{sec:disc}. Our SED modeling in Figure \ref{fig:eazybagsSED} also supports these observations. 
We note that the other sources were not identified by other $z\sim$7-8 studies \citep[e.g.]{Bouwens2015ApJ, rborsani2016ApJ, Stark2017MNRAS}. In fact, these sources have larger $f_{5}/f_{10}$ values compared EGS 29337, which suggest that they appear more point-like, and thus more likely to have been rejected in the galaxy selection. Thus, a unique aspect of this study is that we explore the $z\sim7$-8 dropout point-sources that are often excluded in earlier studies of high-redshift galaxies. 

\subsection{Stellar SED fit properties of point sources}
We list the best-fit stellar population properties of the point sources, which were estimated with \bags\ SED fits, in Table \ref{tab:bagsSED}. Due to the limitation of data, we assume that our detected point sources are
well represented by a young stellar spectrum. The fit results predict sub-solar metallicities for nearly all of our point sources. Considering the young age of the universe
, the metallicity estimates are not surprising. However, since uncertainties in broadband SED fits are strongly influenced by assumptions in the star formation history and age-metallicity-dust degeneracies \citep[e.g., Fig.~12 in][]{Morishita2019ApJ}, it is difficult to place any confident constraints on the metallicity evolution in the early universe. Instead we examine the size and star formation rate (SFR) density properties of the point sources. 

We derive the projected physical size, $R_{\textrm{eff}}$, from the half-light radius, which is calculated with \sex. When compared to the \HSTsurvey\ PSF limit, $0.\!\arcsec14$, 
we find that 2 of the calculated $R_{\textrm{eff}}$ are upper limits. This is similar to the \cite{Morishita2020z8} results, as shown in Figure \ref{fig:LFz8_sizeLum}. This means that these sources are likely unresolved by \hst, despite meeting our point source selection criteria.

Using the SFR estimated from the \bags\ modeling, we also calculate the SFR density, $\Sigma_{\textrm{SFR}}$, which is defined as the average SFR within a circle with radius $R_{\textrm{eff}}$:
\begin{equation}
    \Sigma_{\textrm{SFR}} = \frac{\textrm{SFR} }{ 2\pi R^2_{\text{eff}} }.
    \label{eq:SFRdensity}
\end{equation}
The calculated $\Sigma_{\textrm{SFR}}$ serves as lower limits since its uncertainty is dependent on the upper limit uncertainty in $R_{\textrm{eff}}$. We compare the inferred $\Sigma_{\textrm{SFR}}$, which is calculated from the \muv-SFR relation \citep{Kennicutt1998,Ono2013ApJ} defined as follows:
\begin{equation}
    M_{UV} = -2.5 \log_{10} \bigg[ \frac{ \Sigma_{\textrm{SFR}} \cdot \pi R^2_{\text{eff}} }{2.8\times10^{-28} (M_{\astrosun}yr^{-1})} \bigg]+51.59.
    \label{eq:sigSFR}
\end{equation}
It appears that our point source candidates are highly compact star-forming objects. In Figure \ref{fig:LFz8_sizeLum}, we plot $R_{\textrm{eff}}$ and $\Sigma_{\textrm{SFR}}$ against their corresponding \muv. We also compare the $\Sigma_{\textrm{SFR}}$ redshift evolution. Our results appear to be consistent with trends of high redshift galaxies discussed in \cite{Ono2013ApJ,Holwerda2018AA}, which predicts a greater number of compact galaxies with high star-formation at earlier epochs. 

Using the best-fit \bags\ SEDs, we calculate the UV continuum slope \buv. We adopt the formula defined \citep[e.g.,][]{Dunlop2013MNRAS}:
\begin{equation}
    \beta_{UV} = -2.0+4.39\times(J_{125}-H_{160}),
    \label{eq:Buv}
\end{equation}
where \jj\ and \hh\ here are the best-fit magnitudes from the best-fit \bags\ SEDs. The calculated \buv\ are listed in Table \ref{tab:bagsSED} with the mean slope of $\bar{\beta_{UV}}=-1.90\pm 0.35$. The resulting $\bar{\beta_{UV}}$ is consistent with \buv\ of known bright galaxies at $z\sim$7-8 \citep[e.g.,][]{Dunlop2013MNRAS,Bouwens2014ApJ793}.

Lastly, we estimate the rest-frame equivalent width due to the H$\beta+\oiii$ emission lines from the best-fit SED for objects with sufficient IRAC fluxes. This is possible because $H\beta+\oiii$ emission from $z\sim7$-9 sources is well sampled by IRAC CH1 and CH2, $3.6\,\micron$ and $4.5\,\micron$, \citep{rborsani2016ApJ}. We calculate the equivalent widths as, 
\begin{equation}
    EW_{H\beta+\oiii} = \frac{(f_{\rm ch2}-f_{\rm cont})}{f_{\rm cont}} \frac{\Delta\lambda_{\rm ch2}}{(1+z_{ph})} ,
    \label{eq:EW_hbo3}
\end{equation}
where $f_{\rm cont}$ is the underlying continuum flux obtained from the best-fit \bags\ spectrum, $f_{\rm ch2}$ is the observed \spitz/IRAC CH2 flux, $\Delta\lambda_{\rm ch2}\sim1$\,\micron\ is the full-width half maximum of the CH2 filter, and $z_{ph}$ is from \eazy. Comparing the values in Table \ref{tab:bagsSED} and the SED plots in Figure \ref{fig:eazybagsSED}, we see that the $EW_{H\beta+\oiii}$ estimates can be applied to three targets. These objects show a moderate $EW_{H\beta+\oiii}$ of 500-1000\,\AA, similar to estimates from other high redshift surveys \citep{labbe2013ApJ,Schenker2013ApJ,Smit2014ApJ,Smit2015ApJ,Morishita2020z8}. However, this is based on the assumption that \zph\ is correct.
We compare the redshift evolution of the measured $EW_{H\beta+\oiii}$ in Figure \ref{fig:LFz8_sizeLum}, and our results appear to be consistent with other survey results. 
Future infrared observations with higher resolution and sensitivity are needed to better characterize the predicted $H\beta+\oiii$ emission.





\subsection{Number density of point sources}\label{sec:LF}
From the survey data, we constrain the point source luminosity function at $z\sim7$-8. We produce our own completeness simulation to calculate the effective volume, \veff, probed by the \HSTsurvey. We follow the completeness simulation treatment in \cite{Leethochawalit2021MNRAS} to calculate \veff\ \citep{Oesch2012ApJ745, GLACiAR2018,Calvi2016ApJ817, Morishita2018ApJ}. 

\begin{figure}
	 \begin{center}
    	 \begin{tabular}{c}
    	 \includegraphics[width=0.96\columnwidth]{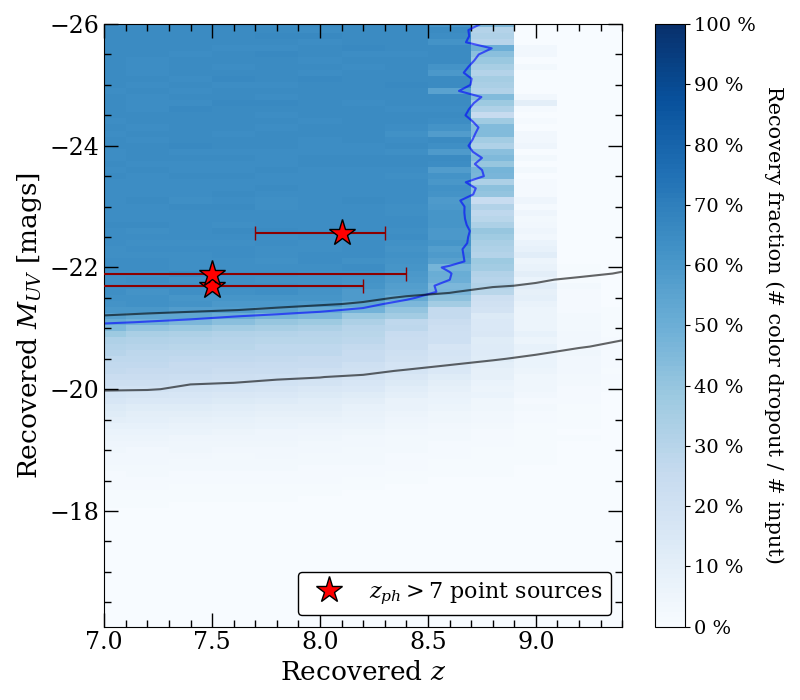}
    	 \end{tabular}
	 \end{center}
	 \caption{The \muv\ and redshift dependent selection probability distribution as determined from our completeness simulation. The plot shows the ratio of color-selected point sources recovered to all input sources as a function of redshift and \muv. The colorbar to the right indicates this recovery fraction. At brighter $M_{UV}\lesssim-21$ magnitudes, at least 50\% of simulated color-selected point sources are recovered (blue contour line). At fainter magnitudes, the recovery fraction decreases due to the combined effect of color-selection and source detection (gray contour lines at 40 \% and 75 \% detection). We also indicate the observed point source candidates at their respective \zph\ and \muv\ (red stars).}
	 \label{fig:compsimul} 
\end{figure}

We inject 500 sources into each \HSTsurvey\ image at each $(M_{UV},z)$ bin: 100 $\Delta M_{UV}$ bins across $-26\leq M_{UV} \leq-16$ and 13 $\Delta z$ bins across $7\leq z \leq9.4$. All simulated sources in a given $(M_{UV},z)$ grid have the same UV slope, which are randomly drawn from a Gaussian distribution with a mean slope of $\bar{\beta_{UV}}=-2.2\pm 0.4$. Source fluxes are calculated in the same way as in \cite{Skelton2014ApJS214}. We extract the simulated point sources according to our selection criteria described in Section \ref{sec:select}. We repeat this process for every field. We show the redshift and magnitude probability distribution function of the extraction completeness in Figure \ref{fig:compsimul}. 

\begin{figure}
	 \begin{center}
    	 \begin{tabular}{c}
    	 \includegraphics[width=0.96\columnwidth]{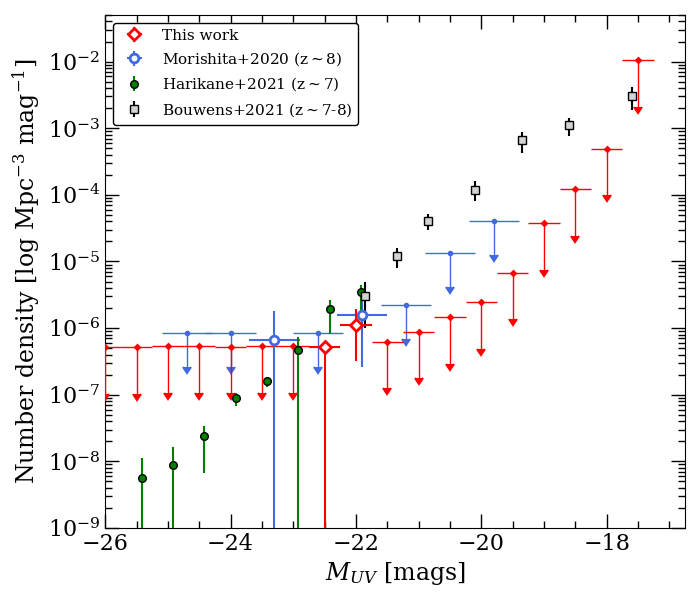} \\
    	 \end{tabular}
	 \end{center}
	 \caption{The derived number density of the \HSTsurvey\ point sources in red, \sborg\ point sources \citep{Morishita2020z8} in blue, and galaxies at $z\sim7$-8 \citep{Bouwens2021AJ} in black. The $z\sim7$ number density from \cite{Harikane2022ApJS} are in green. Open symbols indicate observed data; and filled symbols indicate upper limits, which are estimated from the completeness simulation. Our number density values are listed in Table \ref{tab:LFNdensity}. }
	 \label{fig:numdensity} 
\end{figure}

After we calculate \veff, we estimate the number density of the point sources shown in Table \ref{tab:LFNdensity} within each $\Delta M_{UV}=0.5$ mag bin. We quote Poisson uncertainties for the number density \citep{Gehrels1986ApJ}. In Figure \ref{fig:numdensity}, we plot the estimated number density of our point sources and compare with results from previous surveys of point sources and galaxies at $z\sim7$-8. 

\input{LFNdensity2}

We fit the point source number density with both the Schechter function (Eq.\ref{eq:schechter}; \citealt{Schechter1976ApJ}) and with the Double powerlaw (Eq.\ref{eq:DPL}; \citealt{Hopkins2007ApJ}), where $\phi^*$ is the characteristic normalization, $M^*_{UV}$ is the characteristic UV luminosity defined at \muv\ (1450\,\AA), $\alpha$ defines the faint end slope, and $\beta$ defines the bright end slope. 

\input{LFbestfit2}

\begin{equation}
\begin{split}
    \phi_{\textrm{Sch}} = \frac{\ln{10}}{2.5}\phi^* & \times  10^{-0.4(M'-M_*)(\alpha+1)}\\
    & \times\exp{ \bigg[-10^{-0.4(M_{UV}-M^*_{UV})} \bigg]}
\end{split}
\label{eq:schechter}
\end{equation}

\begin{equation}
\begin{split}
    \phi_{\textrm{DPL}} = \frac{\ln{10}}{2.5}\phi^* \times & \bigg[10^{0.4(\alpha+1)(M_{UV}-M^*_{UV})} \\
    & +10^{0.4(\beta+1)(M_{UV}-M^*_{UV})}  \bigg]
\end{split}
\label{eq:DPL}
\end{equation}

\begin{figure}
	 \begin{center}
    	 \includegraphics[width=\columnwidth]{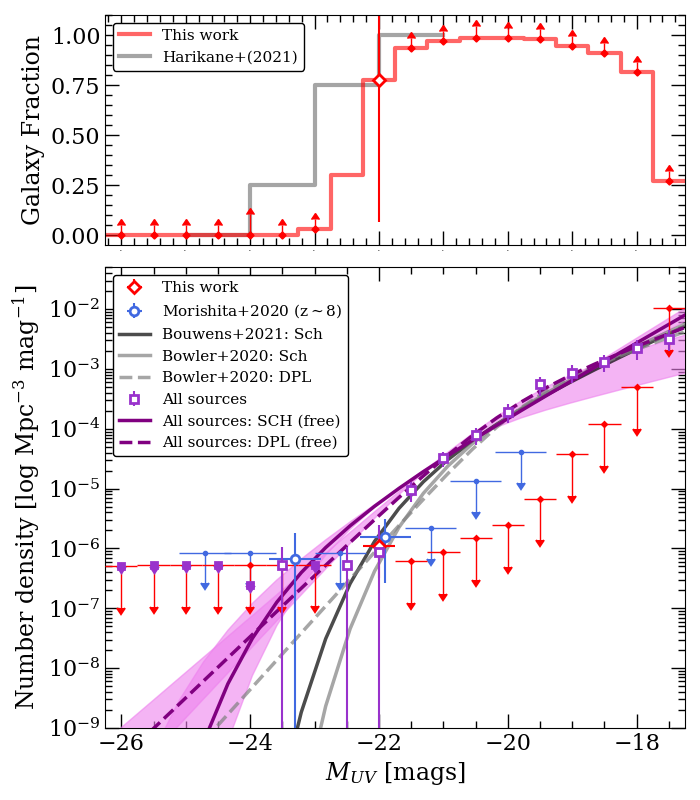}
	 \end{center}
	 \caption{(Top) Comparison of \cite{Bouwens2021AJ} galaxy vs. point source fraction. Point sources dominate a larger fraction of the bright end starting at $\sim-22\textrm{ mags}$. We compare with the \cite{Harikane2022ApJS} galaxy fraction in gray. (Bottom) We compare the best-fit $z\sim7$-8 luminosity functions for the combined, freely-fit point sources and galaxies in purple curves against other galaxy-only models by \cite{Bowler2020MNRAS, Bouwens2021AJ} in black and gray: Schechter in solid, double powerlaw in dashed lines. The purple squares indicate the combined effective number density. EGS 29337 has been removed from the point source number density to avoid duplicate counting. Open symbols indicate observed data; and filled symbols indicate upper limits. The curves are shown with $1\sigma$ uncertainties.}
	 \label{fig:LFz8} 
\end{figure}

\begin{figure*}[!hbt]
	 \begin{center}
    	 \begin{tabular}{cc}
    	 \includegraphics[width=0.9\columnwidth]{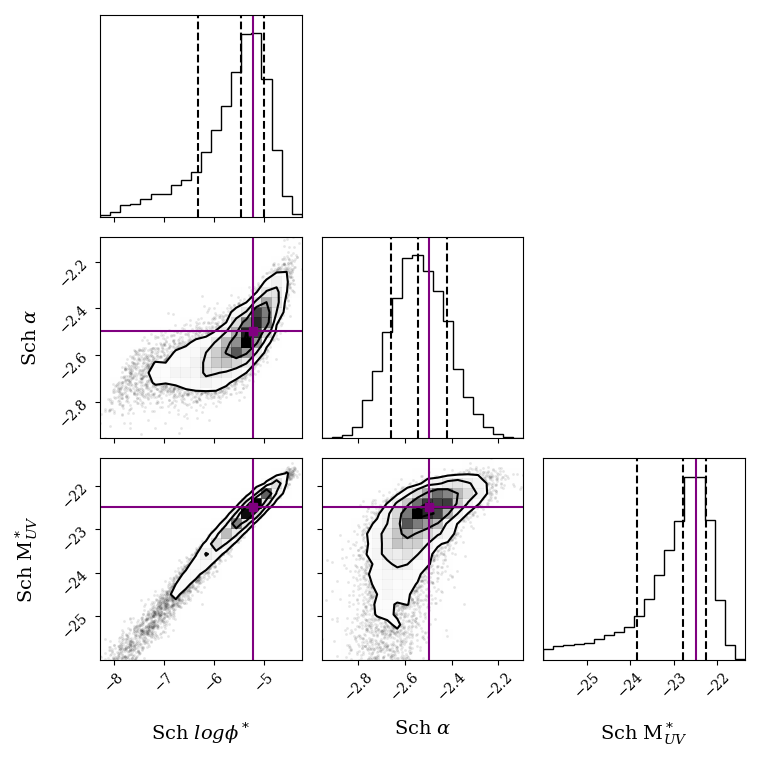} &
    	 \includegraphics[width=1.1\columnwidth]{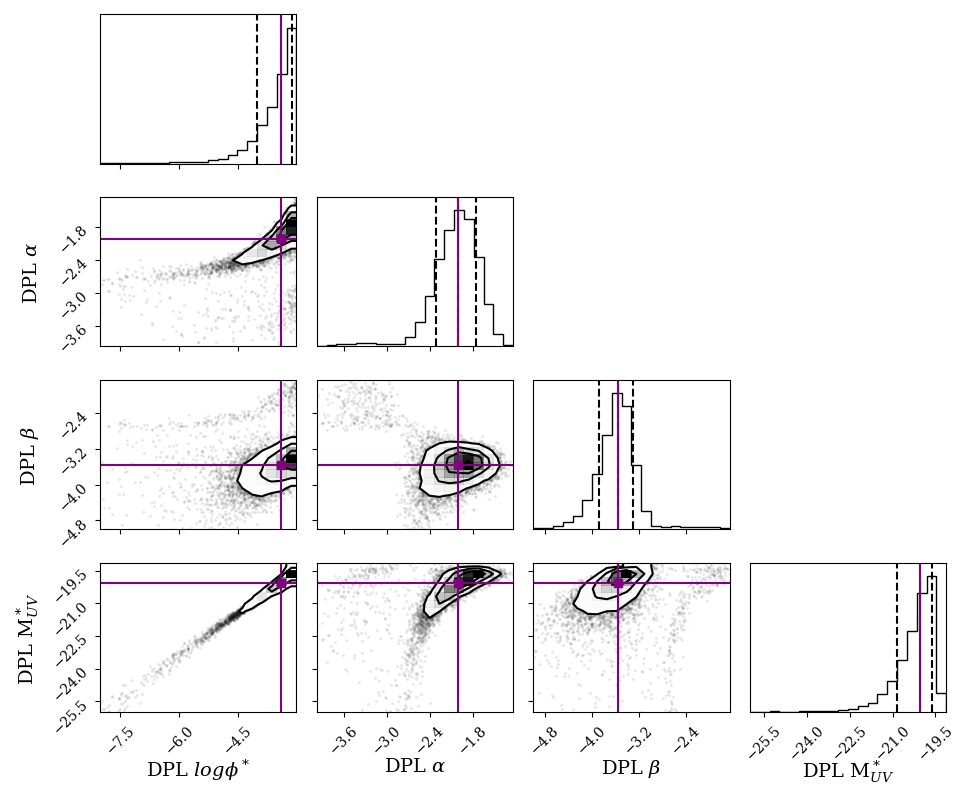}
    	 \end{tabular}
	 \end{center}
	 \caption{Contour plots of the best-fit luminosity function parameters. (Left) Combined point source + galaxy Schechter fits with free parameters. (Right) Combined point source + galaxy double powerlaw fits with free parameters. The best-fit parameters are indicated in Table \ref{tab:LFbestfit}. The 1D histogram of the fit parameters are shown with the 16\%, 50\%, and 84\% percentile values indicated.}
	 \label{fig:LFz8_contour} 
\end{figure*}


To properly include the bins of non-detection of the point source population at the faintest and brightest ends in fitting evaluation, we incorporate the upper limits from non-detections following the derivations from \cite{Sawicki2012PASP} for $\chi^2$ minimization. The derivations for \muv\ and the $\chi^2$ minimization are shown in Appendix \ref{apx:chi2min}. 
We perform Markov Chain Monte Carlo (MCMC) sampling, using \emcee\ package \citep{emcee2013PASP}, to constrain the luminosity function. First we fit the luminosity function for point sources from this study (\HSTsurvey\ selections). Then we apply the fits on all point sources selected from both this study and \sborg. 

If we freely fit for all of the parameters, the fit parameters do not converge to physically meaningful values. This is likely due to the lack of data points at both the faint and bright magnitude ranges. 
Instead, we opt for a more conservative approach and follow the known luminosity function shapes of $z\sim7$-8 galaxies and $z\sim6$ quasars. We fix both the faint end and bright end slopes and freely fit for $\phi^*$ and $M^*_{UV}$. For the Schechter model fits, we fix the faint end slope to $\alpha=-2.2$, which is the observed galaxy luminosity function at $z\sim7$-8 \citep{Bouwens2021AJ}. For the double powerlaw mode fits, we fix both the faint end slope to $\alpha=-1.2$ and the bright end slope to $\beta=-2.7$ based on the extrapolations of the quasar luminosity at $z\sim6$ \citep{Matsuoka2018ApJ869, Harikane2022ApJS}. 

With the deeper exposures of the \HSTsurvey\ survey, we improve the point source luminosity function fits estimated by \cite{Morishita2020z8}. The best-fit luminosity function parameters of the point sources are shown in Table \ref{tab:LFbestfit}. If we freely fit for \muv, we find that only the double powerlaw fit produces more reasonable parameters ($M^*_{UV}\approx-24$) than the best-fit Schechter function, which instead suggests an unrealistically bright UV cut-off ($M^*_{UV}\approx-38$, not shown). This may suggest that the point source luminosity function is more consistent with the high redshift quasar luminosity function (at $z\sim6$; \citealt{Matsuoka2018ApJ869}). On the other hand, if we force a lower-bound on the \muv\ fit, then it is difficult to confidently favor either functions over the other. For both cases, the best-fit normalization $\phi^*$ deviates from the \cite{Matsuoka2018ApJ869} extrapolation by nearly $\times100$. This may be because \HSTsurvey\ is volume-limited, similar to the results in \cite{Morishita2020z8}. 

\subsection{Number density of point source+galaxy populations}
Finally, we fit the combined point source and galaxy luminosity functions. Here, we remove EGS 29337 from the point source luminosity function since it is already included in the \cite{Bouwens2015ApJ} determination (discussed in Section \ref{subsec:res1}). First, we fit with the slopes fixed, and then we freely fit over all parameters. With the current survey volume by \hst, it is difficult to confidently favor either functions over the other for combined luminosity function. With the Schechter model, both the fixed-slope and freely-fit runs produce brighter UV cut-offs at $M^*_{UV}\approx-21.9$, compared to $M^*_{UV}=-22.8$ by \citet{Bouwens2021AJ}, suggesting an excess of bright sources. The freely-fit double powerlaw model produces a superposition of point source Schechter and galaxy Schechter functions also with a slight excess of $M^*_{UV}=-20.0$ and a steep $\beta=-3.6$. 

The exact shape of the galaxy luminosity function is under debate. For example, \cite{Harikane2022ApJS} performed a two-component luminosity function fit (i.e. DPL+DPL or DPL+Schechter) to the combined  quasar+galaxy populations. In contrast, we model a single component function across all magnitudes for both the point source and galaxy number densities (i.e. Schechter or DPL). The main difference between the two studies is that \cite{Harikane2022ApJS} extrapolates the $z\sim6$ \cite{Matsuoka2018ApJ869} quasar luminosity function to estimate the $z\sim7$ relation, while we use observed sources to anchor the point source luminosity function. Despite the differences in modeling, both results demonstrate that the inclusion of bright $M_{UV}\lesssim-24$ \sborg\ sources at $z\sim7$-8 results in a departure from the previously measured galaxy luminosity function, i.e. a bright excess. We plot the $z\sim7$-8 point source number density and the best-fit luminosity functions in Figure \ref{fig:LFz8}. The best-fit parameters of all fits are shown in Table \ref{tab:LFbestfit}. The contours of the combined luminosity function fits are shown in Figure \ref{fig:LFz8_contour}. 

We compare the point source number densities and luminosity function against galaxy luminosity function measured at $z\sim7$-8 \citep{Bouwens2021AJ}. We quantify the fraction of extended galaxies to all sources (galaxies and point sources) as a function of \muv\ as the following:

\begin{equation}
\begin{split}
    f_{galaxy}(M_{UV}) = \frac{\phi_{galaxy}}{\phi_{galaxy}+\phi_{points}}
\end{split}
\label{eq:gfract}
\end{equation}
where $\phi_{galaxy}$ is the galaxy number density from the luminosity function  \citep{Bouwens2021AJ} and $\phi_{points}$ is the observed point source number density listed in Table \ref{tab:LFNdensity}. The uncertainty in $f_{galaxy}$ simply reflects the uncertainty in the number count of point sources (i.e. poisson). We plot $f_{galaxy}$ alongside the luminosity function fits in Figure \ref{fig:LFz8} and compare with results from \cite{Harikane2022ApJS}. We find that these point sources dominate at the bright \muv\ magnitudes. This suggests that the bright end excess implicated by the new point sources is not likely dominated by a typical population that has been identified in previous studies of high-redshift galaxies. We discuss the physical interpretations of this measured excess in the following section. 



\section{Discussion}\label{sec:disc}
Although our sources are selected with slightly different colors and from different surveys, the inferred number density of our point sources at $M_{UV}<-21.5$ mags agree with the \sborg\ point source study. This indicates that these $z\sim7$-8 dropout point sources are abundant enough to be detected in both surveys and are representative of similar populations. The inferred \muv\ suggests that these objects are driven by intense phenomena that occurs in a small physical scale, such as central starburst or quasar activity, which also shape their observed morphology point-like. 
In this section we explore the physical properties and implications of these sources. 

\subsection{Point sources as compact starburst galaxies}
With the exception of EGS 29337, it is currently difficult to distinguish our final candidates between non-active galaxies or quasar-hosting galaxies, the inferred sizes of the point sources is consistent with the observed trends of smaller galaxy sizes in the early universe \citep{Oesch2010ApJ,Ono2013ApJ,Holwerda2015ApJ}. If these point sources are compact non-active galaxies, we predict high $\Sigma_{SFR}$ based on our SED fitting analysis. Previously, \cite{Oesch2010ApJ,Ono2013ApJ} observed constant $\Sigma_{SFR}$ from $z\sim4$ to $z\sim7$ with a weak increase towards higher redshifts. Our SED analysis of the point sources appears to support this increasing $\Sigma_{SFR}$ trend, albeit we predict even larger values, as shown in Figure \ref{fig:LFz8_sizeLum}. Finally, given their predicted SFR and stellar masses, these sources may be progenitors of massive quiescent galaxies that are already present in the early universe \citep[e.g.,][]{vandokkum2008ApJ, Damjanov2009ApJ}. This may suggest that our sources are UV enhanced starbursts and/or that additional physics may be at play. In fact, EGS 29337 has been shown to be one of the brightest $z>7$ known with large SFR \citep{Stark2017MNRAS,rborsani2016ApJ}.



\subsection{Point sources as low-luminosity quasars}
While much of our SED analyses assumed the star-forming SED properties, Figure \ref{fig:eazybagsSED} clearly shows that the SEDs of quasars and star-forming galaxies are degenerate. Theoretical predictions of early quasar properties suggest that variations in the quasar duty cycle may lead to an enhancement of UV bright quasars \citep{RenTrenti2021ApJ}. With the detection of potentially UV bright quasars, there are also implications on the obscured quasar fraction, which is unknown at these redshifts \citep{vito2018MNRAS473, vito2019AA, Inayoshi2020araa}. Either there is an enhancement in the population of unobscured quasars or some physical mechanisms, such as powerful outflows, may drive obscured quasars to appear more luminous like unobscured quasars. Also, while non-detections from deep \textit{Chandra} images suggest that no luminous quasar is present, the possibility of heavy Compton-thick obscuration may complicate this result \citep{Ni2020MNRAS}. Another possibility is that these sources are quasars embedded in star-forming galaxies, similar to $z\sim7$ sources identified by \cite{Laporte2017ApJ}. In fact, the possibility of either a compact starburst or quasar is supported by the recent discovery of an UV compact, red bright, X-ray faint object at $z\sim7$, which is hypothesized to be either a compact dusty, star-forming region or a Compton-thick super-Eddington quasar \citep{Fujimoto2022Nat}.  

Although we cannot distinguish between these possibilities with the current \hst\ data, the \textit{James Webb Space Telescope}'s spatial resolution and sensitivity is expected reach below the predicted sizes and magnitudes of our sources at the range of $ -22 \lesssim M_{UV}\lesssim -18$. \cite{Marshall2021MNRAS} predicts that deep NIRCam observations may allow us to study the quasar and its host galaxy at $z\sim7$. Its imaging and spectroscopic capability may enable us to confidently distinguish them as quasars or as compact star-forming galaxies. If they are revealed as quasars, they will become one of the most distant quasars ever discovered. 

\subsection{Investigating the impact on the bright-end excess}
In this section, we focus on understanding the point sources' contribution to the bright end of the galaxy luminosity function. In Figure \ref{fig:LFz8}, we show the combined point source and galaxy luminosity function. Once the point sources from \HSTsurvey\ and \sborg\ are incorporated, we find that the best-fit luminosity functions suggest the existence of a bright point source population, that may be missed by the galaxy surveys. This may support the existence of a bright end excess in the early universe. While ground-based observations \citep[e.g.]{Bowler2020MNRAS} may detect unresolved sources, these studies are limited to select fields. Thus, we stress the importance of large-volume studies to accurately quantify the luminosity function. 

Indeed, \citealt{Harikane2022ApJS} present a comprehensive analysis of the luminosity function by combining quasar and galaxy populations identified in the HSC program. They proposed several explanations for the apparent bright end excess seen in galaxy luminosity functions. Physical mechanisms such as inefficient mass quenching due to high star-formation and/or poor quasar feedback, low dust obscuration in the host galaxy \citep{MarquesChaves2020, MarquesChaves2021}, or even additional hidden quasar activity may increase the observed rest-frame UV luminosity \citep{Mirocha2020MNRAS}. Variations in the quasar duty cycle can also enhance the UV luminosity and contribute to the bright end \citep{RenTrenti2021ApJ}. Our predictions as compact star-forming galaxies or quasars are consistent with these possibilities.

Other possibilities include superposition of lensed galaxies or even merging galaxies. However, these may be unlikely since the point source criteria requires small elongation values. Our sources also do not appear to be close to potential lensing sources (see Figure \ref{fig:imgTOT}). Moreover, \cite{mason2015ApJ} showed that the effect of magnification bias on the luminosity function determination is small. \cite{Shibuya2022PASJ} calculated the merger fraction of 10\% to 70\% for bright $ -24 \lesssim M_{UV}\lesssim -22$ galaxies \citep{Harikane2022ApJS}. Considering the inferred \muv\ of our sources, there is a non-zero possibility of merger contaminants, especially since galaxy formation in the early universe may involve major mergers.

If the point sources are revealed as quasars by future spectroscopic follow-ups, then it may imply substantial population of low-luminous quasars in the early universe. Since quasar activity is associated with rapid accretion, this may allow a pathway for the rapid formation of massive black hole seeds. Distinguishing between compact sources and contaminants may require higher spatial resolution than the capabilities of \hst. 
What is clear is that point sources selected from deep \HSTsurvey\ and medium-deep \sborg\ both consistently suggest a bright end excess. With the derived upper limit in their number density, we also find that the number of the point-source population dominates only $\simlt5$\,\% over the galaxy population at $M_{\rm UV}\simgt-21$\,mag (Figure \ref{fig:LFz8}). This suggests that their contribution to the faint end luminosity function, and thus cosmic reionization, is likely limited. 

\subsection{Caveat: low-redshift interlopers}\label{sec:contam}
As demonstrated in this paper, the color and morphology selection using current \hst\ capabilities is still challenging in distinguishing between $z\sim7$-8 sources and Galactic interloper stars. The difficulty is further compounded by the fact that these objects are detected at low signal to noise ratio. We also note that the astrometry analysis did not show significant differences in the apparent proper motion following the analysis used in \cite{Morishita2020z8}, so high-quality spectroscopic analysis was the only reliable, albeit incomplete, metric in eliminating low redshift contaminants. We also calculated \buv, defined in Eq.\ref{eq:Buv}, of all SpeX dwarf star template spectra, in the event they are misidentified as galaxies. We find a mean slope of $\bar{\beta_{UV}}=-1.2\pm 1.2$, which is nearly indistinguishable from those of galaxies \citep{Dunlop2013MNRAS}. However, the range of predicted \buv\ is also large, spanning between $-5.32 \lesssim \beta_{UV} \lesssim +6.76$, which suggests that \buv\ is not a useful metric to distinguish galaxies (including quasars) from dwarf stars. In fact, the \buv\ of spectroscopically confirmed dwarf stars (see Appendix \ref{apx:BDlist}) have a mean slope of $\bar{\beta_{UV}}=-4.1$. Thus, we require spectroscopic follow-up to confirm the redshifts and spectroscopic properties of the targets identified in this study. Therefore, for now, our luminosity function estimates serve as an upper limit to the quasar number density at $z\sim7$-8. 

Despite this fact, our study shows that low-resolution spectroscopy around $1\micron$ is an effective method in identifying foreground stars.  
It is noted that when we opt to use the \yy/\jj/\hh\ color selection \citep{Morishita2020z8, morishita2021sb, rborsani2021superB} for the field available (i.e. the GOODS-South field), no dwarf star contaminants are confirmed by the grism data (Ishikawa, in prep.). 
This may be indicative that \yy\ is an effective $z>7$ selector in the absence of other filter observations as proposed by \citet{morishita2021sb}. Despite this challenge, we eliminate a few dwarf stars contaminants with grism spectroscopy. For very faint point sources that do not have sensitive spectroscopic data, we demonstrate that the SED fits favor $z_{ph}\sim7$-8 sources over dwarf stars. 

\section{Conclusion}\label{sec:concl}
We searched for $z\sim8$ Lyman dropout point sources with the archival \HSTsurvey\ data. \HSTsurvey\ surveys nearly 700 square arcminutes of the CANDELS fields, reaching the depth of \jj\ and \hh\ $\sim26$ mags. 
We combined Lyman dropout color and point source selections, with additional photometric redshift estimates and grism spectroscopy to eliminate low-redshift contaminants, and identified three $z\sim7$-8 point source candidates. 

We then investigated the physical properties of the point sources by using the available multi-band photometric data. SED analyses suggest that these sources are potentially quasars or compact star-forming galaxies. Assuming these sources are galaxies, the fitting results revealed high star formation surface density. This is consistent with the redshift trend of previously identified luminous galaxies of comparable luminosity, $M_{UV}\sim-22$; however, we find even larger $\Sigma_{\textrm{SFR}}$ values than those predicted by the relation. We measured the $EW_{H\beta+\oiii}$ emission for the 3 targets with \spitz/IRAC photometry available and found moderately high equivalent widths of $\sim500-1000$\,\AA. 

We calculated the number density distribution and derived the luminosity function of the point sources. 
Similar to previous \hst\ $z\sim7$-8 point source surveys, we found that the inclusion of point sources revealed an excess in the bright end at $M_{UV}\lesssim-22$, consistent with the \sborg\ point source survey. 
We combined the $z\sim7$-8 point source point source candidates with published galaxy number densities to estimate the total galaxy luminosity function. We found that the best-fitting models all point to a bright $M^*_{UV}$ cut-off, which departs from the known galaxy luminosity function. 

The deeper observations of \HSTsurvey\ allowed us to extend the dynamic range covered by our previous work in \sborg. We did not identify point sources in the faint end. 
Moreoever, we found that they make up less than 5\% of the galaxy fraction at the faint end. Thus it is unlikely that these point sources have major contributions to cosmic reionization. 

If these $z\sim7$-8 point sources are confirmed to be quasars, our results suggest that quasars in this luminosity range may be more abundant in the early universe. 
If they turn out to be a galaxy population, it would indicate the presence of compact and intense star-formation in the early universe. Further follow-up observations are required to confirm the inferred properties of our point sources. 
Future surveys using the infrared optimized \textit{James Webb Space Telescope} and large field-of-view capable \textit{Roman Space Telescope} may resolve the current limitations of \hst. 

\acknowledgments

Support for this work was provided by NASA through grant numbers HST-GO-15212.002 and HST-AR-15804.002-A from the Space Telescope Science Institute, which is operated by AURA, Inc., under NASA contract NAS 5-26555.

This work is based on observations taken by the 3D-HST Treasury Program (GO 12177 and 12328) with the NASA/ESA HST, which is operated by the Association of Universities for Research in Astronomy, Inc., under NASA contract NAS5-26555.

This work is based on observations taken by the CANDELS Multi-Cycle Treasury Program with the NASA/ESA HST, which is operated by the Association of Universities for Research in Astronomy, Inc., under NASA contract NAS5-26555.

This research has benefitted from the SpeX Prism Library (and/or SpeX Prism Library Analysis Toolkit), maintained by Adam Burgasser at \url{http://www.browndwarfs.org/spexprism}.

The authors would like to thank the anonymous referee for the helpful suggestions in improving the paper. NL acknowledges that parts of this research were supported by the Australian Research Council Centre of Excellence for All Sky Astrophysics in 3 Dimensions (ASTRO 3D), through project number CE170100013. CAM acknowledges support by the VILLUM FONDEN under grant 37459, the Danish National Research Foundation through grant DNRF140. YI also thanks Nadia Zakamska for helpful discussions.

\software{\astropy\ \citep{Astropy2013},  
        \eazy\ \citep{Brammer2008ApJ686},
        \bags\ \citep{Carnall2018},
        \sex\ \citep{Bertin1996},
        \lmfit\ \citep{lmfit2014zndo},
        \emcee\ \citep{emcee2013PASP},
        \texttt{xspec} \citep{xspec1996},
        \texttt{CIAO} \citep{ciao2006SPIE}}



\bibliography{z8ps_3dhst_apj_v2}
\bibliographystyle{aasjournal}


\input{appendix_apj_v2}



\end{document}

%% file: targlist.tex
\begin{table*}[!ht]
    \caption{Target selected from the \HSTsurvey\ fields based on color and morphology criteria. Photometric redshifts estimated with \eazy. The targets listed here are predicted to have photometric redshift probability of $p(z_{ph}>6)>70  \%$. We also indicate whether grism spectra is available in the \HSTsurvey\ database. Fluxes are listed in Table \ref{tab:photom}. We also show the \eazy\ $\chi_{\nu}$ goodness-of-fit for the \zph\ and dwarf star SED fits. }
    \label{tab:targlist}
    \begin{center}
    \begin{tabular}{lcccccccccc}
    \hline \hline
    Target & RA & DEC & $e$ & $f_5/f_{10}$ & $I_{814} - J_{125}$  & $J_{125} - H_{160}$ & \zph & $p(z_{ph}>6)$ &  $\chi_{\nu,z_{ph}}^2$ & $\chi_{\nu,BD}^2$ \\
           & (J2000)  & (J2000) & $(a/b)$  &  -  & (ABmag) & (ABmag) & - & - &-&-\\
    \hline															
GDS 45797   & $53.036877$ & $-27.696156$ & 1.008 & $0.98\pm0.01$ & $>3.4$ & $0.31\pm0.08$   & $8.1^{+0.3}_{-0.4}$ &	99 \%    & $4.67$ & $9.42$ \\

EGS 515	    & $215.252975$ & $+53.028187$ & 1.111 & $0.60\pm0.02$	& $>2.2$ & $-0.40\pm0.16$  & $7.5^{+0.7}_{-0.6}$ &	99  \%  & $2.54$ & $9.70$ \\
EGS 29337   & $215.050369$ & $+53.007496$ & 1.024 & $0.53\pm0.02$ & $>2.4$ & $0.02\pm0.09$   & $7.5^{+0.9}_{-0.8}$ &	99  \%  & $1.61$ & $12.45$ \\
    \hline		
    \end{tabular}
    \end{center}
\end{table*}

%% file: targphotom.tex
\begin{table*}[!ht]
    \caption{\hst\ and \spitz/IRAC aperture photometry taken at $0.\!''7$ apertures from the \HSTsurvey\ catalog. Only these values are used for the color dropout selection. Non-detections for both \hst\ and \spitz\ are shown at $2\sigma$ upper limits. The uncertainties of real detections are defined at $1\sigma$.  
    } 
    \label{tab:photom}
    \begin{center}
    \begin{tabular}{lcccccccccc}
    \hline \hline
    Target  & F435W  & F606W & F775W   & F814W  & F105W  & F125W   & F140W   & F160W   & 
    CH1     & CH2    
    \\
  & ($\mu$Jy) & ($\mu$Jy)   & ($\mu$Jy)& ($\mu$Jy)   & ($\mu$Jy)   & ($\mu$Jy)   & ($\mu$Jy) & ($\mu$Jy)   &
  ($\mu$Jy)  & ($\mu$Jy) 
  \\
    \hline
    GDS 45797 & $<0.04$& $<0.04$ & $<0.02$ & $<0.02$ & - & $0.40\pm0.03$ & - & $0.54\pm0.01$ &
    -$^{\ddagger}$ & -$^{\ddagger}$ 
    \\
    EGS 515  & - & $<0.04$ & - & $<0.04$ & - & $0.28\pm0.02$ & - & $0.19\pm0.02$ & $0.10\pm0.06$ &
    $0.38\pm0.06$
    \\
    EGS 29337& - & $<0.06$ & - & $<0.03$ & $0.168\pm0.05^{\dagger}$ & $0.30\pm0.01$ & - & $0.30\pm0.02$ & $0.43\pm0.02$ & 
    $1.00\pm0.08$ 
    \\
    \hline
    \end{tabular}
    \end{center}
    \tablecomments{$^{\dagger}$ The \yy\ flux for EGS 29337 is taken from the $Y_{105} - J_{125}$ value in \cite{rborsani2016ApJ}.\\$^{\ddagger}$\spitz/IRAC fluxes with visually confirmed contamination are excluded.}
\end{table*}

%% file: BAGpriors.tex
\begin{table}[!pt]
    \caption{\bags\ SED fit priors assuming a young stellar population model. The model redshifts are fixed to the \eazy-derived \zph\ found in Table \ref{tab:targlist}. $\tau_{age}$ is the range of universe ages calculated with the exponential star-formation history model; $M_{\star}$ is the final stellar mass formed, $Z$ is the metallicity, $A_{V}$ measures the dust attenuation, and $U$ is the nebular ionization parameter that captures the nebular emission and continuum components.}
    \label{tab:BAGpriors}
    \begin{center}
    \begin{tabular}{ccc}
    \hline \hline
    Parameter & Units & Priors  \\
    \hline
    $z$          & - & \eazy\ \zph  \\
    $\tau_{age}$ & (Gyr) &[0.001,1]  \\
    $\log_{10}(M_{\star}/M_{\astrosun})$ & - & [6,15]  \\
    $Z$ & $(Z_{\astrosun})$ &[0,1]  \\
    $A_{V}$ & mags &[0,1]  \\
    $\log_{10}U$ & - &$-3$  \\
    \hline
    \end{tabular}
    \end{center}
\end{table}

%% file: BAGbestfit.tex
\begin{table*}
    \caption{Best-fit \bags\ SED parameters for each object. The model redshifts are fixed to the \eazy\ \zph\ in Table \ref{tab:targlist}. The model priors are listed in Table \ref{tab:BAGpriors}. PSF-limited $R_{\rm eff}$ are shown as upper limits.}
    \label{tab:bagsSED}
    \begin{center}
    \begin{tabular}{lcccccccccc}
    \hline \hline
    Target & $R_{\rm eff}$ & \muv & \buv &SFR & $\tau_{\rm age}$ & $\log_{10}(M_*/M_{\astrosun})$ & $\log_{10}(Z/Z_{\astrosun})$ & $\Sigma_{\rm SFR}$ & $EW_{H\beta+\oiii}$\\
          & (kpc) & (ABmags)&- & (\myr) & (Gyr) & - & - & (\myr kpc$^{-2}$) & (\AA) \\    
    \hline															
GDS 45797 & $<0.7$ & $-22.56$ & $-2.10$ & $12\pm7$ & $0.5\pm0.3$ & $9.1\pm0.2$  & $-1.70\pm0.43$  & $>34$ & -   \\
EGS 515   & $0.8\pm0.7$ & $-21.86$ & $-2.20$ & $15\pm7$ & $0.6\pm0.3$ & $9.2\pm0.3$  & $-0.57\pm0.43$ & $>9$   & $1200\pm500$ \\
EGS 29337 & $0.8\pm0.7$ & $-21.85$& $-1.40$  & $46\pm27$ & $0.6\pm0.2$ & $10.3\pm0.2$ & $-0.60\pm0.42$ & $>11$  & $500\pm200$  \\
    \hline		
    \end{tabular}
    \end{center}
\end{table*}

%% file: xraytable.tex
\begin{table}[!pt]
    \caption{Upper limits on the \textit{Chandra} X-ray rest-frame fluxes and luminosities. We assume a simple powerlaw with $\Gamma=2$ at \zph\ without any obscuration.}
    \label{tab:xray}
    \begin{center}
    \begin{tabular}{lcc}
    \hline \hline
    Target & $F_{\textrm{0.5-2 keV}}$ & $L_{\textrm{2-10 keV}}$\\
     & ($\textrm{erg cm}^{-2} \textrm{ s}^{-1}$) & ($\textrm{erg s}^{-1}$)\\
    \hline															
GDS 45797  & $<4.3\times10^{-17}$ & $<4.3\times10^{43}$ \\
EGS 515	   & $<8.0\times10^{-17}$ & $<6.2\times10^{43}$ \\
EGS 29337  & $<4.0\times10^{-17}$ & $<3.1\times10^{43}$ \\
    \hline		
    \end{tabular}
    \end{center}
\end{table}

%% file: LFNdensity2.tex
\begin{table}[!tp]
    \caption{Number density, $\Phi$, of 3D-HST point-sources at $z\sim8$. \muv\ is the rest-frame UV luminosity at 1450\AA\ obtained from the best-fit \bags\ SED. The \veff\ is calculated by the completeness simulation \citep{Leethochawalit2021MNRAS}. The $1\sigma$ uncertainties in $\Phi$ is based on \cite{Gehrels1986ApJ}. We also show the $1\sigma$ upper limits for $\Phi$, where appropriate. We take $\Delta$\muv$\approx0.5$ mag bins to match \cite{Bouwens2021AJ}.}
    \label{tab:LFNdensity}
    \begin{center}
    \begin{tabular}{cccc}
    \hline \hline
    \muv & \veff & \# obj & $\Phi$ \\
    (ABmag) & ($10^3$ Mpc$^3$) &  & (Mpc$^{-3}$ mag$^{-1}$) \\
    \hline
    $-26.0$	&	$1941$	&	0	&	$<5\times10^{-7}$	\\
    $-25.5$	&	$1918$	&	0	&	$<5\times10^{-7}$	\\
    $-25.0$	&	$1870$	&	0	&	$<5\times10^{-7}$	\\
    $-24.5$	&	$1884$	&	0	&	$<5\times10^{-7}$	\\
    $-24.0$	&	$1900$	&	0	&	$<5\times10^{-7}$	\\
    $-23.5$	&	$1885$	&	0	&	$<5\times10^{-7}$	\\
    $-23.0$	&	$1864$	&	0	&	$<5\times10^{-7}$	\\
    $-22.5$	&	$1910$	&	1	&	$<5\times10^{-7}$	\\
    $-22.0$	&	$1790$	&	2	&	$(11\pm8)\times10^{-7}$	\\
    $-21.5$	&	$1611$	&	0	&	$<6\times10^{-7}$	\\
    $-21.0$	&	$1139$	&	0	&	$<9\times10^{-7}$	\\
    $-20.5$	&	$680$	&	0	&	$<1\times10^{-6}$	\\
    $-20.0$	&	$410$	&	0	&	$<2\times10^{-6}$	\\
    $-19.5$	&	$146$	&	0	&	$<7\times10^{-6}$	\\
    $-19.0$	&	$26$	&	0	&	$<4\times10^{-5}$	\\
    $-18.5$	&	$8$	    &	0	&	$<1\times10^{-4}$	\\
    $-18.0$	&	$2$	    &	0	&	$<5\times10^{-4}$	\\
    $-17.5$	&	$0.1$	&	0	&	$<1\times10^{-2}$	\\

    \hline
    \end{tabular}
    \end{center}
\end{table}

%% file: LFbestfit2.tex
\begin{table*}[!th]
    \caption{Best fit luminosity function parameters. For the point sources, we fix both the $\alpha$ and $\beta$ slopes depending on the model used and freely fit for $\phi^*$ and $M^*_{UV}$. Parameters that are fixed are shown in a square bracket. The Schechter model slope is fixed to the galaxy faint-end slope from \cite{Bouwens2021AJ}, and the double power-law slopes are fixed to the AGN function slopes based on $z\sim6$ \citep{Matsuoka2018ApJ869,Harikane2022ApJS}. We also fit for a combined galaxy and point source model with the slopes fixed and with freely fit parameters. The fit contours corresponding to the errors are shown in Figure \ref{fig:LFz8_contour}.
    }
    \label{tab:LFbestfit}
    \begin{center}
    \begin{tabular}{ccccccc}
    \hline \hline
    Survey & Model & $\phi^*$ & $M^*_{UV}$ & $\alpha$ & $\beta$  \\
    & & (Mpc$^{-3}$ mag$^{-1}$) & (ABmag) & &  \\
    \hline
Point sources: this work 
& Schechter & 
$(1.5^{+4.3}_{-1.0})\times10^{-8}$ &$ -24.8^{+1.1}_{-0.9}$ &  $[-2.2]$ & $-$  \\
& Double Power Law & 
$(3.7^{+4.3}_{-1.7})\times10^{-7}$ & $-23.5^{+1.2}_{-1.4}$ & $[-1.2]$ & $[-2.7]$  \\
Point sources: this work + \sborg 
& Schechter & 
$(1.0^{+3.4}_{-0.7})\times10^{-8}$ &$-24.7^{+1.2}_{-0.9}$ & $[-2.2]$ & $-$  \\
& Double Power Law & 
$(2.1^{+3.3}_{-1.1})\times10^{-7}$ & $-23.6^{+1.5}_{-1.5}$ & $[-1.2]$ & $[-2.7]$  \\
Point sources +
Galaxies  (fixed $\alpha$, $\beta$)
& Schechter & 
$(2.7^{+1.0}_{-0.8})\times10^{-5}$  & $-21.9^{+0.2}_{-0.2}$ & $[-2.2]$ & $-$   \\
& Double Power Law & 
$(7.7^{+0.4}_{-0.3})\times10^{-3}$  & $-17.6^{+0.2}_{-0.2}$ & $[-1.2]$ & $[-2.7]$  \\
Points sources +
Galaxies (freely fit)  
& Schechter & 
$(3.4^{+6.3}_{-2.9})\times10^{-6}$ & $-22.8^{+0.5}_{-1.0}$ & $-2.5^{+0.1}_{-0.1}$ & $-$  \\
& Double Power Law & 
$(4.1^{+3.8}_{-3.1})\times10^{-4}$ & $-20.0^{+0.4}_{-0.8}$ & $-2.0^{+0.3}_{-0.3}$ & $-3.6^{+0.3}_{-0.3}$  \\
    \hline
    \end{tabular}
    \end{center}
\end{table*} 

%% file: appendix_apj_v2.tex
\appendix 
\section{Other point sources of interest}\label{apx:sourceX}
Here we list the 3 non-candidate point sources that meet the selection criteria (color, shape, and \zph), but have large \ii\ uncertainties and land outside of the selection box. A deeper \ii\ image will determine their \ii$-$\jj\ color. 
\input{targX2}

\section{Composite quasar spectrum}\label{apx:qsotemp}
\begin{figure*}[!th]
	 \begin{center}
    	 \includegraphics[width=0.7\textwidth]{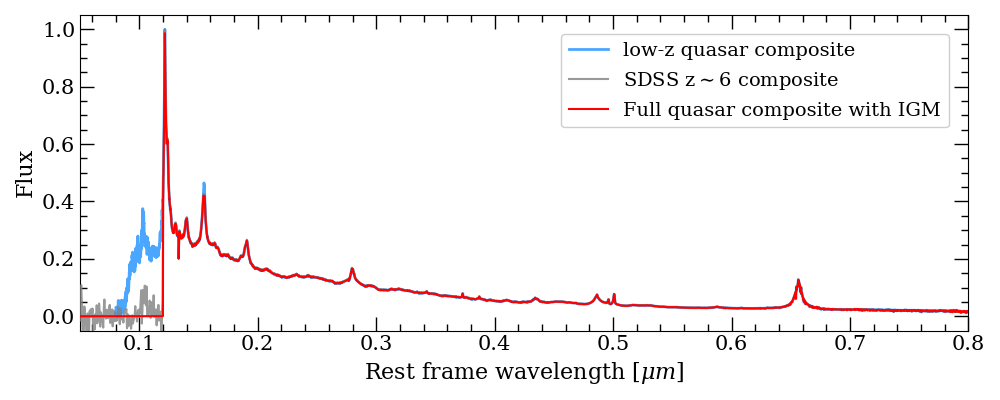} 
	 \end{center}
	 \caption{Low-redshift template (blue), $z\sim6$ SDSS composite (gray), and full composite quasar spectrum (red).}
	 \label{fig:qsotemp} 
\end{figure*}

We generate a composite quasar spectrum for the SED fits in Sec.~\ref{sec:photz}. We combine the low redshift templates  \citep{vandenberk2001,Glikman2006ApJ} and $z\sim6$ spectra from SDSS \citep{sdssQ14}. Since the neutral hydrogen absorption due to the intergalactic medium is unknown at $z\sim7$-8, we assume full absorption at $\lambda<1215$ \AA. We do not adjust the emission lines such as: \ly, which will be affected by intergalactic absorption, \civ, which is expected to show a blueshifted profile \citep[e.g.,][]{yang2021ApJ}, or \hb\ $+$ \oiii. As we can seen in Figure \ref{fig:eazybagsSED}, our composite quasar spectrum lacks the strong nebular emission lines, which are predicted by the \bags\ star-forming galaxy SEDs. Detailed spectroscopic analysis is beyond the scope of this study, and will be left for future study. The final composite quasar spectrum is shown in Figure \ref{fig:qsotemp}.

\section{Dwarf star selections}\label{apx:BDlist}

\input{BDlist}

\begin{figure*}[!th]
	 \begin{center}
    	 \begin{tabular}{cc}
    	 \includegraphics[width=0.45\textwidth]{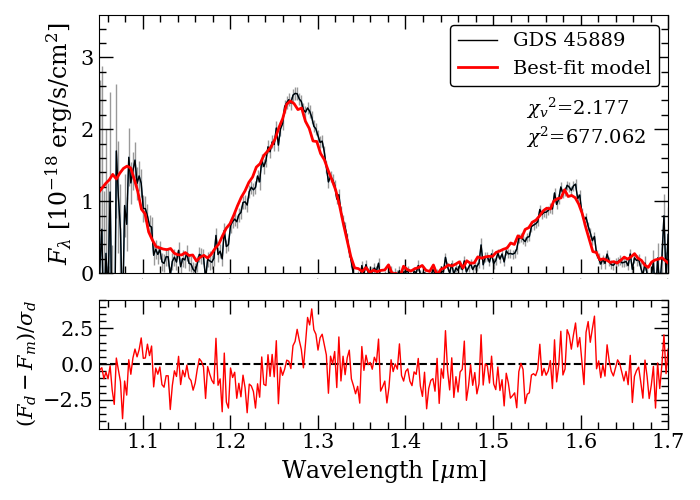} &
    	 \includegraphics[width=0.45\textwidth]{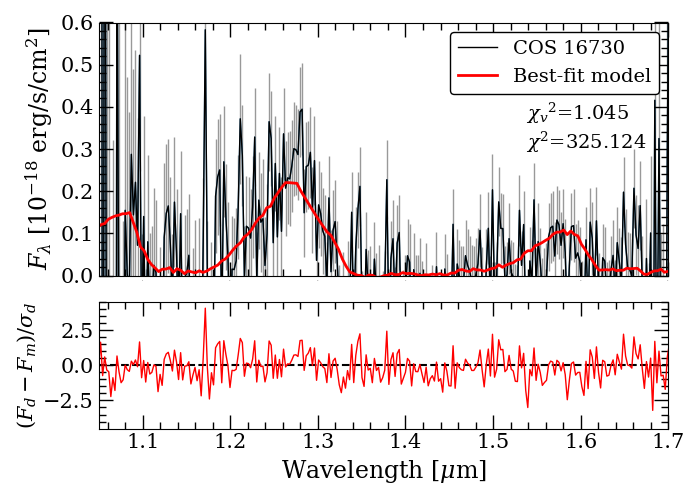} \\
    	 \includegraphics[width=0.45\textwidth]{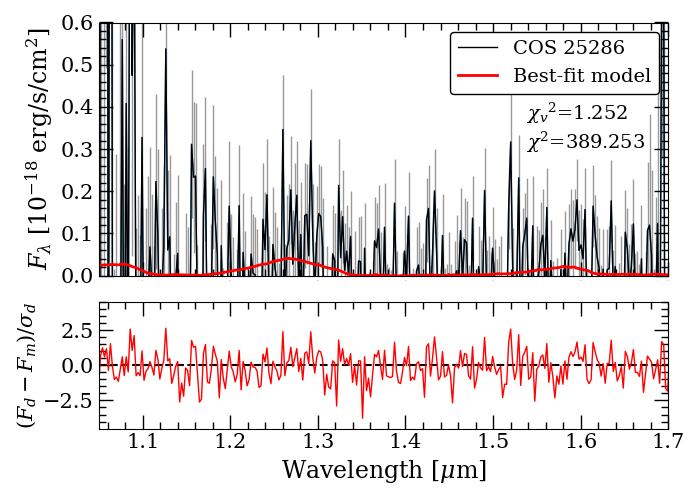} &
    	 \includegraphics[width=0.45\textwidth]{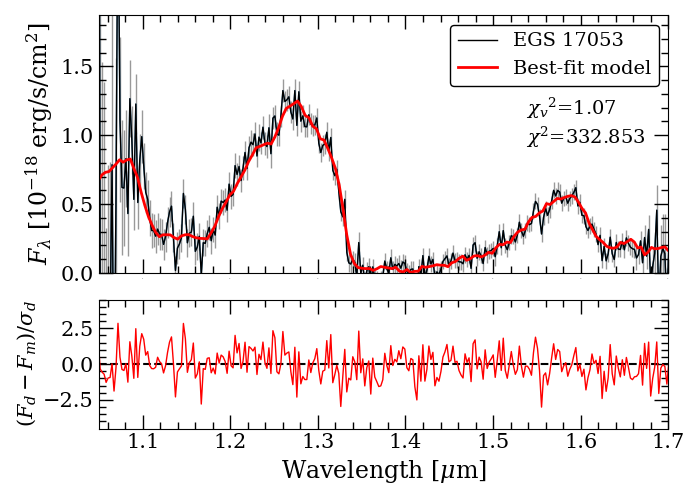} \\
    	 \end{tabular}
	 \end{center}
	 \caption{The observed \HSTsurvey\ grism spectra in black with the best-fit SpeX templates in red.}
	 \label{fig:grismBD} 
\end{figure*}

Due to the nature of point source selection, it is inevitable that we find dwarf star contaminants as discussed in Section \ref{subsec:lowz}. From the short list of color dropout point sources, we identify 4 dwarf stars (GDS 45889, COS 16730, COS 25286, and EGS 17053) based on \HSTsurvey\ grism spectroscopic data. We perform a least-squares fit, minimizing the residual, $(F_d-F_m)/\sigma$. $F_d$ is the observed grism flux; $F_m$ is the scaled template flux; and $\sigma_d$ is the uncertainty of the grism flux. The sources are best fit with T7, T7.5, T8, and T5.5 templates respectively to within $3\sigma$. We note that the best fitting templates are approximate and are only used to identify dwarf star contaminants. Accurate characterization will require careful stellar modeling, as demonstrated by \cite{Aganze2021}. With the exception of COS 25286, all of these sources were identified by \cite{Aganze2021}; it is possible that COS 25286 did not make their selection due to its low $S/N$ ratio. These targets have F125W magnitudes ranging between $25\gtrsim I_{125}\gtrsim 22$ mags. While three out of four dwarf stars are much brighter than our final targets, some like COS 25286 can appear as faint as $z\sim8$ candidates. If we calculate the \buv\ of these targets using Eq.\ref{eq:Buv}, we find very blue slopes of $\bar{\beta_{UV}}=-4.1\pm 0.5$. We show the target fluxes in Table \ref{tab:BDflux}; the grism spectra and their best fitting SpeX templates \citep{Rayner2003PASP,spexprism2014} are shown in Figure \ref{fig:grismBD}. 

\section{Fitting the luminosity function}\label{apx:chi2min}
We calculate the absolute rest-frame UV magnitude, \muv, at \zph\ as follows: 
\begin{equation}
     M_{UV} = m_{UV/(1+z_{ph})} - 25 - 5\log_{10}\bigg[(3 \textrm{ Mpc}^{-1})\times(1+z_{ph}) \times DA(z_{ph}) - 5\log_{10}h \bigg]+2.5\log_{10}(1+z_{ph}),
\label{eq:muv}
\end{equation}
where $h$ is the Hubble parameter and $DA(z_{ph})$ is the angular diameter distance in Mpc, calculated with \astropy.
To fully reflect the estimated number densities, including upper limits, 
we fit the luminosity function by minimizing Eq.~\ref{eq:chi2} as derived by \cite{Sawicki2012PASP}, which is also re-written in a more convenient form for computation:
\begin{equation}
\begin{split}
     \chi^2_{mod} & = \sum_{i} \bigg(\frac{\phi_{d,i}-\phi_{m,i}}{\sigma_i}\bigg)^2 
      - 2\sum_{j}\ln\int_{-\infty}^{\phi_{\ulim,j}}\exp \bigg[-\frac{1}{2}\bigg(\frac{\phi'-\phi_{m,i}}{\sigma_j} \bigg) \bigg]d\phi' \\
      & = \sum_{i} \bigg(\frac{\phi_{d,i}-\phi_{m,i}}{\sigma_i}\bigg)^2 
      - 2\sum_{j}\ln \bigg\{ \sqrt{\frac{\pi}{2}}\sigma_j \bigg[1+\erf\bigg(\frac{\phi_{ulim,j}-\phi_{m,j}}{\sigma_j} \bigg) \bigg]\bigg\},
\end{split}
\label{eq:chi2}
\end{equation}
where $\erf(x)=(s/\sqrt{\pi})\int_0^x e^{-t^2}dt$. Here $\phi_{d,i}$ is the observed number density at a given $\Delta$\muv\ bin; $\phi_{m,i}$ is the model luminosity function value at the same \muv\ bin; $\phi_{\ulim,j}$ is the upper limit number density; and $\sigma$ is the uncertainty in observed number density. When detections are made at all bands, the second summation ($j$-index) goes to zero, revealing the standard $\chi^2$ form.

%% file: targX2.tex
\begin{table*}[!h]
    \caption{Basic parameters of the \ii\ $S/N$ limited targets.}
    \label{tab:targlistX2}
    \begin{center}
    \begin{tabular}{lcccccccccc}
    \hline \hline
    Target & RA & DEC & $e$ & $f_5/f_{10}$ & $I_{814} - J_{125}$  & $J_{125} - H_{160}$ & \zph & $p(z_{ph}>6)$ &  $\chi_{z_{ph}}^2$ & $\chi_{BD}^2$ \\
           & (J2000)  & (J2000) & $(a/b)$  &  -  & (ABmag) & (ABmag) & - & - &-&- \\ 
    \hline															
GDS 29369   & $53.104618$ & $-27.776581$ & 1.015 & $0.64\pm0.003$ & $>4.9$ & $-0.35\pm0.03$  & $7.1^{+0.1}_{-0.1}$ &	100  \%   &	$8.76$ & $27.69$ \\
GDS 41852   & $53.029492$ & $-27.714358$ & 1.127 & $0.55\pm0.03$ & $>2.16$ & $0.12\pm0.2$   & $7.7^{+0.7}_{-5.9}$ &	71  \%    &0.82&4.43	\\
EGS 6572    & $214.965836$ & $+52.855282$ & 1.061 & $0.50\pm0.04$ & $>1.27$ & $-0.04\pm0.23$  & $7.5^{+0.9}_{-1.0}$ &	90  \%  &1.65&6.17  \\
EGS 13621   & $214.901611$ & $+52.839909$ & 1.189 & $0.57\pm0.05$ & $>1.07$ & $0.04\pm0.26$   & $7.0^{+1.0}_{-0.9}$ &	91  \% &10.31&9.00  \\
    \hline		
    \end{tabular}
    \end{center}
\end{table*}

%% file: BDlist.tex
\begin{table}[!h]
    \centering
    \caption{The \hst\ and \spitz/IRAC aperture photometry taken at $0.\!''7$ apertures of the spectroscopically confirmed dwarf stars. These targets meet the color dropout and point source morphological selections. Non-detections for both \hst\ and \spitz\ are shown at $2\sigma$ upper limits. The uncertainties of real detections are defined at $1\sigma$. The best-fit \hst\ grism spectra, which also correspond to fluxes in F140W, are shown in Figure \ref{fig:grismBD}.}
    \label{tab:BDflux}
    \begin{tabular}{lccccccccccccc}
    \hline \hline
    Target & RA & DEC &  F814W   & F125W   & F140W   & F160W   & CH1     & CH2  \\
           & (J2000)  & (J2000)  & ($\mu$Jy) & ($\mu$Jy) & ($\mu$Jy) & ($\mu$Jy) & ($\mu$Jy) & ($\mu$Jy) \\
    \hline
    GDS 45889 & $53.242542$	& $-27.695446$  & $0.10\pm0.05$ & $4.48\pm0.02$ &	$5.02\pm0.04$ &	$2.43\pm0.02$ &	$2.66\pm0.10$ &	$6.09\pm0.09$ \\
    
    COS 16730 &	$150.178040$ &	$+2.349691$	&	$<0.02$ &	$0.55\pm0.02$ &	$0.57\pm0.04$ &	$0.38\pm0.02$ &	$0.28\pm0.07$ &	$1.02\pm0.08$ \\
   
    COS 25286	&	$150.137115$	&	$+2.440077$	&	$<0.03$ & $0.33\pm0.02$ &	$0.23\pm0.05$ &	$0.22\pm0.02$ &	$0.15\pm0.06$ &	$0.14\pm0.06$ \\
    
    EGS 17053	&	$214.710007$	&	$+52.716480$	&	$0.07\pm0.03$ &	$2.93\pm0.02$ &	$3.00\pm0.03$ &	$2.06\pm0.03$ &	$1.93\pm0.06$ &	$3.68\pm0.06$ \\
    \hline															
    \end{tabular}
\end{table}